\documentclass[onecolumn,apj,numberedappendix,twocolappendix]{openjournal}

\usepackage[utf8]{inputenc}
\usepackage{txfonts}
\usepackage{wrapfig}
\usepackage{amssymb}

\usepackage{graphicx}
\DeclareGraphicsExtensions{.bmp,.png,.jpg,.pdf}
\usepackage{float}
\usepackage{multirow}
\usepackage{makecell}
\usepackage{float}

\usepackage{xcolor}
\definecolor{linkcolor}{rgb}{0.0,0.3,0.5}
\usepackage[table]{xcolor} 
\usepackage{soul}
\usepackage{color,colortbl}

\usepackage{textgreek}
\usepackage[autopunct=true,autostyle=false]{csquotes}

\usepackage{tensind}
\tensordelimiter{?}
\usepackage{verbatim}
\usepackage[normalem]{ulem}
\usepackage{orcidlink}
\usepackage{soul}
\usepackage{tabularx}
\newcommand{\orcidauthor}[3]{\author{\href{http://orcid.org/#1}{#2$^{#3}$}}}
\DeclareUnicodeCharacter{0306}{}

\defcitealias{Castellano2024}{C24}

\newcommand{\UVLFrefs}{Adams2024, Adams2025, Asada2026, Bouwens2023a, Bouwens2023b, Casey2024, Castellano2025Z17, Chemerynska2026, Donnan2023, Donnan2024, Finkelstein2023, Finkelstein2024, Franco2025, Gandolfi2026b,  Harikane2024, Harikane2025, Kokorev2025, Kreilgaard2026, Leung2023, McLeod2024, McLeod2026, Moreira_Goolsby2026, PerezGonzalez2023, PerezGonzalez2025, Robertson2024, Rojas-Ruiz2025, Weibel2026UVLF, Whitler2025, Willott2024}

\newcommand{\UVLFfig}{Adams2024, Adams2025, Asada2026, Bouwens2023a, Bouwens2023b, Casey2024, Castellano2025Z17, Chemerynska2026, Donnan2023, Donnan2024, Finkelstein2023, Finkelstein2024, Franco2025, Gandolfi2026b,  Harikane2024, Harikane2025, Kokorev2025, Kreilgaard2026, Leung2023, McLeod2024, McLeod2026, Moreira_Goolsby2026, PerezGonzalez2023, PerezGonzalez2025, Robertson2024, Rojas-Ruiz2025, Weibel2026UVLF, Whitler2025, Willott2024}

\newcommand{\UVLFfit}{Adams2024, Adams2025, Asada2026, Bouwens2023a, Bouwens2023b, Casey2024, Castellano2025Z17, Chemerynska2026, Donnan2023, Donnan2024, Finkelstein2023, Finkelstein2024, Franco2025, Gandolfi2026b,  Kokorev2025, Kreilgaard2026, Leung2023, McLeod2024, McLeod2026, Moreira_Goolsby2026, PerezGonzalez2023, PerezGonzalez2025, Robertson2024, Rojas-Ruiz2025, Weibel2026UVLF, Whitler2025, Willott2024}

\usepackage{hyperref}
\hypersetup{
    unicode,
    colorlinks=true,
    linkcolor=linkcolor,
    citecolor=linkcolor,
    filecolor=linkcolor,
    urlcolor=linkcolor,
}

\shorttitle{}
\shortauthors{A. Fontana et al.}
\catcode`\'=12

\begin{document}
\title{Too many or too bright? Status and perspectives in the study
of the UV Luminosity Function at $z\geq10$\vspace{-1.5cm}}

\orcidauthor{0000-0003-3820-2823}{A. Fontana$^\dagger$}{1}
\orcidauthor{0000-0002-7570-0824}{H. Atek}{2}
\orcidauthor{0000-0003-0212-2979}{V. Bromm}{3,4}
\orcidauthor{0000-0002-8651-9879}{A.~J.\ Bunker}{5} 
\orcidauthor{0000-0001-9875-8263}{M. Castellano}{1}
\orcidauthor{0000-0001-8460-1564}{P. Dayal}{6,7,8}
\orcidauthor{0000-0003-2388-8172}{F. D'Eugenio}{9,10}
\orcidauthor{0000-0002-7631-647X}{D. Elbaz}{11}
\orcidauthor{0000-0002-9400-7312}{A. Ferrara}{12,13}
\orcidauthor{0000-0001-8519-1130}{S.~L. Finkelstein}{3,4}
\orcidauthor{0000-0001-7201-5066}{S. Fujimoto}{7}
\orcidauthor{0000-0002-7831-8751}{M. Giavalisco}{14}
\orcidauthor{0000-0002-6047-430X}{Y. Harikane}{15}
\orcidauthor{0000-0002-4343-0487}{C. Kobayashi}{16}
\orcidauthor{0000-0001-9879-7780}{F. Pacucci}{17}
\orcidauthor{0000-0003-4528-5639}{P.~G. P\'erez-Gonz\'alez}{18}
\orcidauthor{0000-0002-6813-0632}{P. Rosati}{19}
\orcidauthor{0000-0002-1899-4360}{A. Trinca}{20}
\orcidauthor{0000-0002-9909-3491}{R. Tripodi}{1}
\orcidauthor{0000-0002-9334-8705}{P. Santini}{1}
\orcidauthor{0000-0002-5057-135X}{E. Vanzella}{21}
\orcidauthor{0000-0003-2237-0777}{A. Venditti$^*$}{3,4}
\orcidauthor{0000-0001-8928-4465}{A. Weibel}{22}

\affiliation{$^1$INAF Osservatorio Astronomico di Roma, Via Frascati 33, 00078 Monteporzio Catone, Rome, Italy}
\affiliation{$^{2}$Institut d'Astrophysique de Paris, CNRS, Sorbonne Universit\'e, 98bis Boulevard Arago, 75014, Paris, France}
\affiliation{$^{3}$Department of Astronomy, University of Texas at Austin, 2515 Speedway, Stop C1400, Austin, TX 78712, USA}
\affiliation{$^{4}$Cosmic Frontier Center, The University of Texas at Austin, Austin, TX 78712}
\affiliation{$^5$Department of Physics, University of Oxford, Denys Wilkinson Building, Keble Road, Oxford OX1\,3RH, U.K.}
\affiliation{$^6$Canadian Institute for Theoretical Astrophysics, 60 St George St, University of Toronto, Toronto, ON M5S 3H8, Canada}
\affiliation{$^7$David A. Dunlap Department of Astronomy and Astrophysics, University of Toronto, 50 St George St, Toronto ON M5S 3H4, Canada}
\affiliation{$^8$Department of Physics, 60 St George St, University of Toronto, Toronto, ON M5S 3H8, Canada}
\affiliation{$^9$ Kavli Institute for Cosmology, University of Cambridge, Madingley Road, Cambridge, CB3 0HA, United Kingdom}
\affiliation{$^10$ Cavendish Laboratory - Astrophysics Group, University of Cambridge, 19 JJ Thomson Avenue, Cambridge, CB3 0HE, United Kingdom}
\affiliation{$^{11}$Universit\'e Paris-Saclay, Universit\'e Paris Cit\'e, CEA, CNRS, AIM, 91191, Gif-sur-Yvette, France}
\affiliation{$^12$Scuola Normale Superiore,  Piazza dei Cavalieri 7, Pisa I-56126, Italy}
\affiliation{$^13$Centro B. Segre, Accademia Nazionale dei Lincei, Via della Lungara, 10, Roma I-00165, Italy}
\affiliation{$^{14}$University of Massachusetts Amherst, 710 North Pleasant Street, Amherst, MA 01003-9305, USA}
\affiliation{$^{15}$Institute for Cosmic Ray Research, The University of Tokyo, 5-1-5 Kashiwanoha, Kashiwa, Chiba 277-8582, Japan}
\affiliation{$^{16}$University of Hertfordshire, College Lane Campus, Hatfield, Hertfordshire, AL10 9AB, UK}
 \affiliation{$^{17}$Center for Astrophysics $\vert$ Harvard \& Smithsonian, 60 Garden St, Cambridge, MA 02138, USA}
\affiliation{$^{18}$Centro de Astrobiolog\'{\i}a (CAB), CSIC-INTA, Ctra. de Ajalvir km 4, Torrej\'on de Ardoz, E-28850, Madrid, Spain}
\affiliation{$^{19}$Dipartimento di Fisica e Scienze della Terra, Universit\`a degli Studi di Ferrara, Via Giuseppe Saragat 1, 44122 Ferrara, Italy}
\affiliation{$^{20}$Institute for Astronomy, University of Edinburgh, Royal Observatory, Blackford Hill, Edinburgh, UK}
\affiliation{$^{21}$INAF - OAS, Osservatorio di Astrofisica e Scienza dello Spazio di 
Bologna, via Gobetti 93/3, I-40129 Bologna, Italy}
\affiliation{$^{22}$ Institute for Astronomy, University of Hawai‘i, 2680 Woodlawn Drive, Honolulu, HI 96822, USA}

\thanks{\vspace{0.1cm}$^\dagger$E-mail: \href{mailto:adriano.fontana@inaf.it}{adriano.fontana@inaf.it}}
\thanks{\vspace{0.1cm}$^*$Cosmic Frontier Center Prize Fellow}
\begin{abstract}

The finding that  galaxies in the first 500\,Myr were much more numerous and/or brighter than expected is often cited as one of the main discoveries of the James Webb Space Telescope in its first four years of operation.  This result is now solidly based on an impressive number of surveys that have detected, spectroscopically confirmed, and characterized a relatively large sample of galaxies up to $z=14.44$, the redshift of the most distant galaxy currently known. This finding has motivated the introduction of additional physical mechanisms into theoretical models to reproduce the observed numbers and luminosities. These results contrast with the failure, despite ongoing efforts, to confirm any galaxy candidate beyond this redshift.
This review summarizes the status of the search for and characterization of galaxies at $z>9$, focusing on the evolution of the UV Luminosity Function (UVLF), which is the main statistical description of the galaxy population that  can be obtained at these redshifts. We first provide an comprehensive  overview of the full set of JWST data obtained so far. We perform a \enquote{meta--analysis} on the entire data set and provide a simple yet effective description of the  global evolution of the UVLF, averaging over the various surveys. We also compare these results extensively with theoretical models that were developed to match the new JWST observations and that explore different physical mechanisms able to drive the observed evolution. We conclude with a summary of the contrasting results regarding the search for galaxy candidates at $z>15$, and outline the requirements for future surveys aimed at breaking this barrier.

\end{abstract}

\maketitle

\section{Introduction}\label{sec:intro}

Few facilities in the recent history of astronomy have transformed our understanding of a research field as profoundly and as quickly as the James  Webb Space Telescope (JWST) has transformed the search for and  characterization of the earliest galaxies.

Within the first weeks after the first data arrived in July 2022, {\it a)} a surprisingly large number of bright galaxies at $z>10$ were discovered \citep{Castellano2022b, Naidu2022b, Finkelstein2022Maise}, far exceeding the numbers expected from previous HST studies; {\it b)} a new class of objects, the so-called Little Red Dots (LRDs), was found, initially identified as candidate massive galaxies by \citet{Labbe2023} and later recognized as a separate class \citep[e.g.,][]{Matthee2024,Greene2024}. Within a few months, {\it c)} the record for the most distant galaxy with a spectroscopic redshift was raised to $z=13.2$ \citep{Curtis-Lake2022}, and {\it d)} objects with extreme spectral properties were identified at $z>10$ \citep{Bunker2023}.

These early discoveries were quickly confirmed and expanded in the following years. A solid body of observational evidence 
on the evolution of galaxies at $z>10$ is now in place, both in  terms of their number densities \citep[see, e.g.,][for the latest results; full references are given below]{McLeod2026,Weibel2026UVLF} and in terms of their physical properties \citep[e.g.,][and references therein, and Sect.~\ref{sec:disc_prop}]{Roberts-Borsani2026}. There is increasing evidence that some of these early galaxies host  Super Massive Black Holes (SMBH) \citep[e.g.,][]{Larson2023, maiolino2024, Napolitano2024b}: these low-luminosity Active Galactic Nuclei (AGNs) may contribute to the UV luminosity and ionizing photon budget and affect the nature of these early galaxies \citep{Dayal2025b}. 
Some of these BHs may have masses so large as to challenge the standard paradigm of BH formation through Eddington-limited growth of stellar-mass seeds, leading to more exotic proposals such as super-Eddington growth, heavy astrophysical seeds (direct-collapse BHs, runaway collisions), or even primordial BHs \citep{Liu_PBH2022, dayal2024_pbh, ziparo2025, prole2025, dayal2026, Zhang2026BH}.

Finally, tantalizing early results on the possible detection of Population~III (Pop~III) star-forming clusters have emerged \citep{Matteri2025, Jeon_HEBE2026, Maiolino_HEBE2026, Ubler2026, Venditti2026}.



Spectroscopic follow-up of photometrically selected LRDs has now revealed a remarkably heterogeneous population, spanning large ranges in continuum shape, Balmer-break strength, emission-line properties, gas density, stellar properties, and inferred BH activity \citep[e.g.,][]{perezgonzalez2026}. Most show broad Balmer lines indicative of massive or supermassive BHs \citep[e.g.,][]{wang2024a, kokorev2024LRD, Juodzbalis2024}, whereas several canonical AGN signatures, such as hard X-rays from hot coronae, near- to mid-infrared emission from hot dust, and radio jets, are weak or absent \citep{maiolino2025, yue2024, akins2024, perez2024, casey2025}. Rather than converging on a single picture, proposed models include dust-enshrouded AGNs, compact starbursts, tidal disruption events, BH stars, and binary massive BHs \citep{inayoshimaiolino2025, inayoshi2025a, inayoshi2025b, inayoshi2025c, kido2025, bellovary2025, perger2025, Naidu2026, rusakov2025, DeGraaff2026, Madau2026}, but none yet provides a comprehensive explanation. Despite a lively debate about their structure and evolution, LRDs remain poorly understood, and many suspect that they hold key information about the earliest evolutionary phases of the galaxy-AGN system, including the interplay between Pop~III stars and supermassive BH seeds, the early growth of these seeds, and the formation of the host galaxy.

This wealth of results is not unexpected: the telescope's imaging and spectroscopic capabilities were designed with these scientific goals in mind.
Four years after the first data release, time is ripe to reassess the status of these fields, consolidate the many discoveries into a coherent picture and identify the open challenges and fundamental questions that remain unresolved.

Focusing  on the evolution of galaxies at $z>10$, the spectacular progress of the past four years raises several questions that remain open. What drives the unexpectedly high efficiency of galaxy assembly in the first 500~Myr after the Big Bang? Which physical mechanisms produced the diversity in size, chemical enrichment, and ionization state observed in these objects? Are these galaxies caught at different stages of a common evolutionary path, or do they follow independent evolutionary sequences that eventually lead to different classes of objects? 
The most intriguing question is why we have not (yet) confirmed any galaxy beyond $z\sim 14.5$ (the current record holder is MoM-z14 at $z=14.44$; \citealt{Naidu2025}), despite the presence of relatively bright galaxies up to this redshift and the high success rate of spectroscopic confirmation for candidates  up to this redshift. Is this a purely observational limitation, due to the small number of spectroscopic follow-up observations and the faintness of the targets, or does it signal a drastic change in galaxy properties?

\begin{wrapfigure}{l}{0.5\textwidth}
\centering
\includegraphics[trim={0.25cm 0.25cm 0.25cm 0.1cm},clip,width=0.95\linewidth,keepaspectratio]{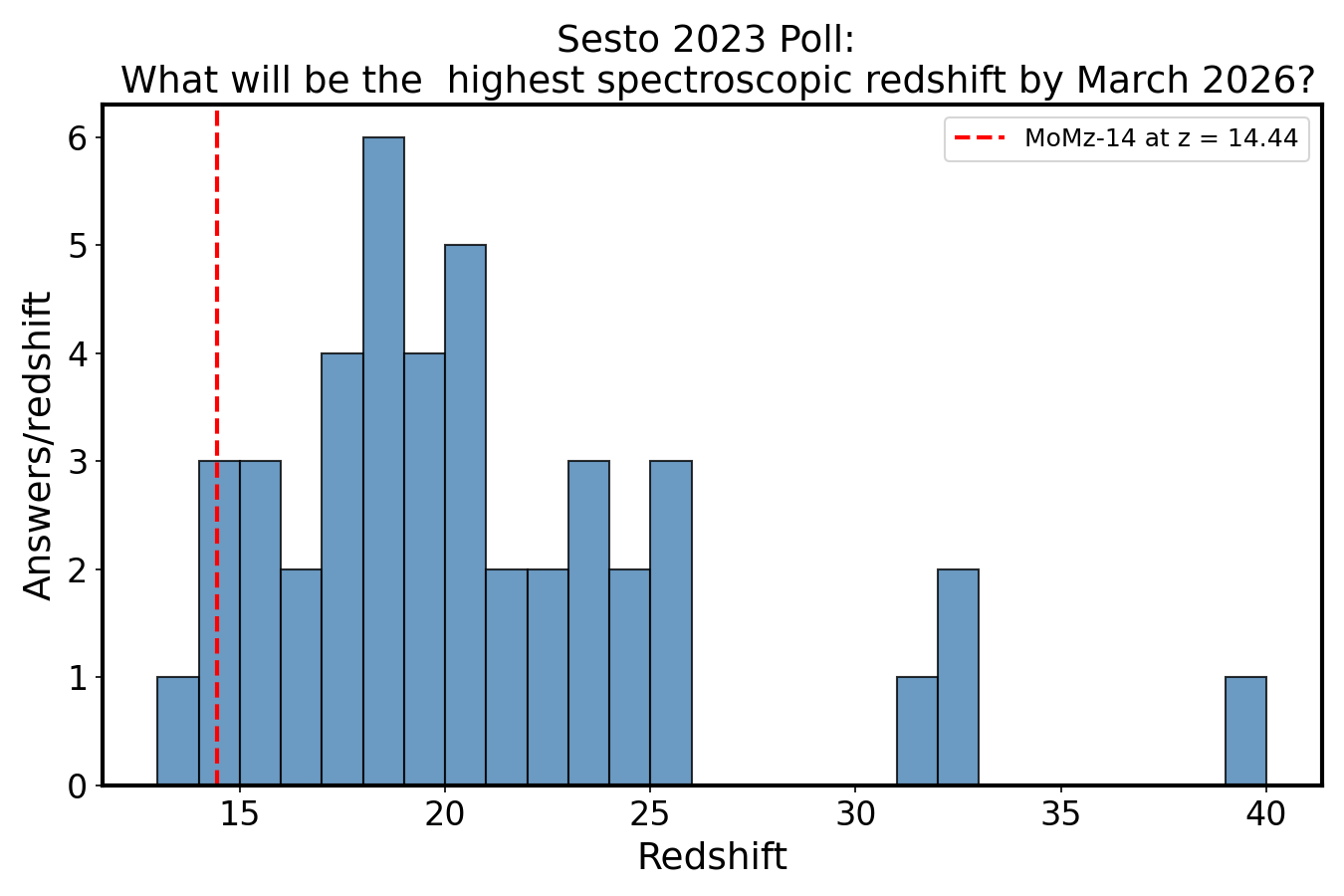}
\caption{Distribution of the responses to the question \enquote{What will be the redshift of the highest-$z$ confirmed galaxy three years from now?} posed to the participant of the meeting "The Growth of Galaxies in the Early Universe -- VIII: the Physics of Early Galaxies" held in Sesto, Italy, in March 2023. The red vertical line shows the position of MoM-z14 \citep{Naidu2025}, the highest-redshift confirmed galaxy both at the deadline (March 2026) and at the time of writing }
\label{fig:SestoBet}
\end{wrapfigure}

An anecdote may be useful to put these questions in context. At a meeting held in Sesto in March 2023, less than one year after the first data release, the participants, impressed by the extraordinary results obtained in such a short time, including the spectroscopic confirmation of objects up to $z=13.2$ \citep{Curtis-Lake2022}, were asked in a provocative poll to bet on \enquote{What will be the redshift of the highest-redshift confirmed galaxy three years from now?}. 
Now that the March 2026 deadline has passed, it is instructive to compare these predictions with the current status (Fig.~\ref{fig:SestoBet}). 
Leaving aside the obvious jokes about the ability of professional astronomers to predict the evolution of their own field, the optimism of most participants conveys the excitement that pervaded the scientific community in the early months of the JWST revolution and explains why this question is felt so keenly today.

In this context, the meeting \textit{CSI: Sesto -- Cosmology Scene Investigation in Sesto: Joining Efforts to Solve JWST Mysteries in the Distant Universe} organized at the Sexten Center for Astrophysics ``Riccardo Giacconi'' (Sesto, Italy) on January 26--30, 2026,  brought together experts to discuss three topics in particular: \textit{a)} the observational and theoretical efforts to discriminate between the scenarios proposed to explain the excess of bright galaxies at $z>10$; \textit{b)} strategies to achieve the long-awaited discovery of the first generation of stars (Pop~III), and \textit{c)} efforts to unveil the  nature of LRDs and their role in galaxy evolution. 

This review summarizes and extends the outcome of the lively discussions on the first topic, deliberately focusing on one aspect: the description and interpretation of the evolution of the galaxy UV Luminosity Function (UVLF). We refer the reader to \citet{Venditti2026} for an extended review of  the searches and theoretical expectations  of Pop~III stars, also discussed at that meeting.

\section{The evolution of the UV Luminosity Function at \texorpdfstring{$z\geq10$}{z>10}}\label{sec:UVLF}

\subsection{Crossing the boundary at  \texorpdfstring{$z\sim10$}{z~10}: the transition from HST to JWST}\label{sec:LF10}

We begin by reexamining, after four years of JWST results, the basis for the claim that galaxy evolution at $z>10$ is ``accelerated'' relative to the predictions drawn from pre-JWST surveys.  As we show below, these predictions were driven by the final HST results, based on the complete datasets of the ultra-deep surveys conducted with HST in several deep fields. For simplicity, we use the UVLF as the primary statistical tracer of galaxy evolution.

Tracing the evolution of the UVLF from $z\sim4$ to $z\sim9$ is arguably one of the major legacies of two decades of HST and ground-based surveys. Several of these programs \citep[see e.g.][and \citealt{Finkelstein2022CANDELS} and references therein]{Adams2020,Atek2018,Bagley2022,Bouwens2017,Bouwens2021,Bouwens2022,Bowler2014,Bowler2020,Castellano2010,Castellano2010b,Finkelstein2015,Harikane2022,Ishigaki2018,McLeod2016,Morishita2018,Parsa2016,Stefanon2019}
described the overall evolution of galaxy properties across this broad redshift range. The final comprehensive analyses revealed an approximate plateau between $z=2$ and $z=4$, followed by a smooth decline of the UVLF at $z>4$ \citep[see, e.g., Fig.~5 of][]{Bouwens2021}, in line with the expectations of most theoretical models. The breadth and quality of the data enabled a global description of the UVLF in terms of a smoothly evolving analytic function (either a Schechter function or a double power law, hereafter DPL; \citealt{Bouwens2021,Finkelstein2022AGNGAL}).  At $z\sim10$, however, the capabilities of HST (and, even more so, of ground-based instruments) were limited not only in depth but especially in wavelength coverage. With the Lyman break moving into the $J_{125}$ filter at $z>9$, the selection of $z\sim10$ galaxies relied only on simultaneous detections in the $JH_{140}$ and $H_{160}$ bands (to exclude spurious sources and transients), a large $J_{125}-JH_{140}$ color break, and a blue continuum redward of the break, constrained by Spitzer/IRAC data up to $4.5\,\mu$m that were shallow and had poor spatial resolution.

\begin{wrapfigure}{r}{0.5\textwidth}
\centering
\includegraphics[trim={0.25cm 0.25cm 0.25cm 0.1cm},clip,width=0.95\linewidth,keepaspectratio]{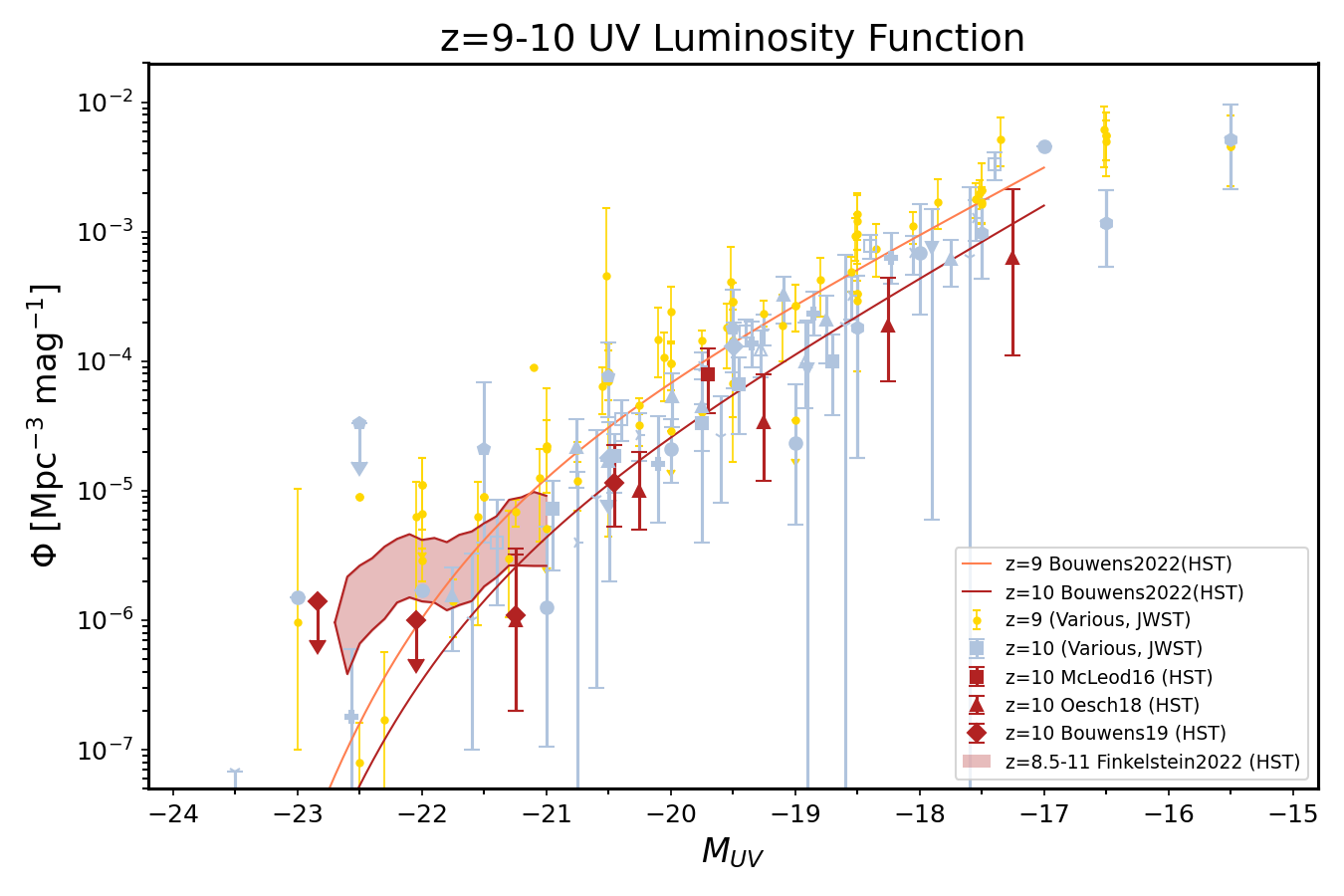}
\caption{Measurements of the UV luminosity function at $z=9$ and $z=10$ from all JWST surveys conducted so far (yellow and light blue, respectively), compared with the final HST-based measurements at $z=10$ (red points and shaded area; \citealt{McLeod2016,Oesch2018,Bouwens2019,Finkelstein2022CANDELS}). The orange (red) line shows the \citet{Bouwens2021} fit to the UVLF at $z=10$). The JWST measurements at $z=9$--10 are the same as those used in Fig.~\ref{fig:uvlfz9to14}.}
\label{fig:uvlfz09to10}
\end{wrapfigure} 

After a series of initial explorations \citep{Ellis2013,McLure2013,Oesch2014,Ishigaki2015,Bouwens2015}, the final analysis of the legacy HST surveys yielded a small sample of $z\sim10$ candidates: two from \citet{McLeod2016}, nine from \citet{Oesch2018}, and four from \citet{Finkelstein2022CANDELS} (with some overlap among the samples), plus a bright $z\simeq 10$ candidate selected in ground-based VISTA images \citep{Bowler2020}, later shown spectroscopically to be a lower-redshift interloper \citep{Harikane2025}. Unsurprisingly, the picture emerging from these surveys was dominated by large statistical fluctuations, leading to discrepancies among the results.  The best known example is the tension between the  evolution of the UV Luminosity Density (UVLD) from $z\sim9$ to $z\sim10$. Measurements by \cite{McLeod2016}  appeared to be consistent with the steady decline observed at $z>6$, contrary to the apparently faster decline observed by \cite{Oesch2018}. This scatter is visible in Fig.~\ref{fig:uvlfz09to10}, where these surveys are plotted in red.

When the largest $z\sim10$ sample, that of \citet{Oesch2018}, was combined with the more robust samples at $z=6$--9, the data indicated strong evolution beyond $z\sim8$, as shown by \cite{Bouwens2021} and \cite{Finkelstein2022AGNGAL}. In the former, the authors performed a global fit to the UVLF to derive a smooth analytic description of the redshift evolution of the Schechter parameters. This fit indicated significant evolution at $z>8$ that, when extrapolated to $z>10$, played a major role in setting low expectations for the JWST number counts. It is instructive, however, to note how large the uncertainties in the Schechter fit were at $z=10$ \citep[see Fig.~6 of][]{Bouwens2021}. A similar analysis up to $z\sim9$ \citep{Finkelstein2022AGNGAL} also shows uncertainties increasing toward the highest redshifts. Wide-area pure-parallel surveys also pointed to a slightly higher number density at $z\sim9$ than in the legacy fields \citep{Morishita2018,Rojas-Ruiz2020, Rojas-Ruiz2025}. In Sect.~\ref{sec:theory} we  discuss the role of theoretical models, tailored to match the HST results, in predicting the number of galaxies at $z>10$.

The many surveys conducted in the four years since the first JWST data release now make it possible to reexamine the robustness of the HST results and their role in predicting the overall evolution of the UVLF. In Fig.~\ref{fig:uvlfz09to10} we plot the UVLFs measured with JWST at $z=9$ and $z=10$, compared with the HST measurements at $z=10$ \citep{McLeod2016,Oesch2018,Bouwens2019,Finkelstein2022CANDELS}. The JWST data clearly confirm the evolution from $z=9$ to $z=10$, but this evolution is less pronounced than that implied by the \citet{Oesch2018} UVLF. The same conclusion follows from a comparison with the HST-based fits of \citet{Bouwens2021}: at $z=9$ (orange line) the fit agrees with the JWST data (yellow points), whereas at $z=10$ (red line) it lies somewhat below the bulk of the JWST data (light blue points), especially at $M_{\rm UV}\lesssim-20$.

Notably, the early HST results, in particular those of \citet{Oesch2018}, were not biased high, as one might expect if low-S/N interlopers had contaminated the selection; rather, they underestimated the true number densities. Moreover, most of these early HST candidates were later confirmed by JWST imaging and spectroscopy, with the notable exception of the  VISTA candidate mentioned above.
This indicates that the stringent selection criteria adopted by \citet{McLeod2016,Oesch2018,Bouwens2019,Finkelstein2022CANDELS} were sound, if anything too conservative, and that the large uncertainties and the overall underestimate may stem from completeness corrections or  cosmic variance at the bright end, occurring as these works tried to push HST and Spitzer to their limits. This is a lesson to keep in mind when interpreting the JWST results at the current redshift frontier, as we discuss below.

\subsection{Into the JWST realm: the evolution of the UVLF at \texorpdfstring{$9<z<14.5$}{9<z<14}}\label{sec:LF12}

\begin{figure}[ht]
\centering
\includegraphics[trim={2cm 4cm 2cm 4cm},width=0.95\linewidth,keepaspectratio]{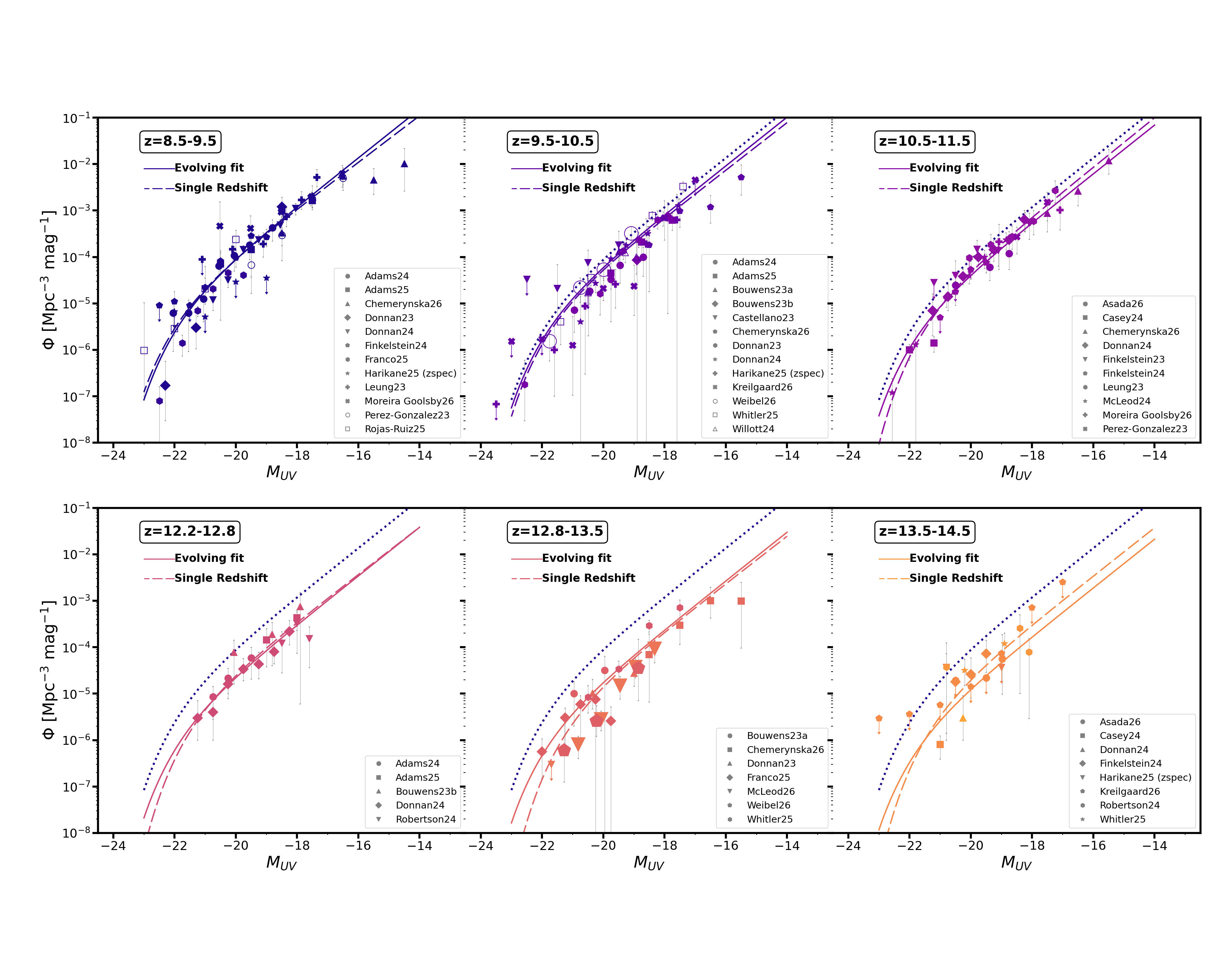}
\caption{Measurements of the UV luminosity function from the JWST surveys conducted so far, from $z\sim8.5$ (upper left panel) to $z\sim14.5$ (lower right panel); the redshift interval of each panel is given in the legend. Data are from \citet{\UVLFfig}. The solid lines show our fiducial evolving Schechter model, in which the three parameters $\alpha(z)$, $\log\phi^*(z)$, and $M^*_{\rm UV}(z)$ evolve as in Eq.:~\ref{Eq:EvolvingSchechter}. The solid line shows  the final "Fiducial" fit with only 4 free parameters in total. The dashed lines show independent Schechter fits to the data in each redshift interval. In all panels, the dotted line shows our $z=9$ UVLF fit for comparison.}
\label{fig:uvlfz9to14}
\end{figure}

In four years of JWST operation, many studies have remeasured the UVLF at and beyond $z\sim10$ \citep{\UVLFrefs}. 
These studies build on a large body of data from progressively deeper and wider NIRCam surveys. Compared with the earlier HST$+$Spitzer data, the superior resolution, depth, and wavelength coverage of NIRCam imaging have led to a much more robust selection of candidates.These selections have been validated by a growing number of NIRSpec spectroscopic confirmations. Imaging surveys have reached depths of $m_{\rm AB}\simeq31.5$ with exposure times exceeding 100\,ks \citep{Leung2023,PerezGonzalez2025} and, thanks to deep observations of lensing fields, absolute magnitudes as faint as $M_{\rm UV}\simeq-15$ to $-14$ \citep{Chemerynska2026,Moreira_Goolsby2026}. The spectroscopic confirmation of $z>10$ candidates has been highly successful up $z\sim 14.5$ \citep[e.g.,][]{Curtis-Lake2022,ArrabalHaro2023, ArrabalHaro2023Nat, Carniani2024a, Fujimoto2023b, Hsiao2023, Hsiao2024, Naidu2025, Williams2023, Castellano2024, Roberts-Borsani2023, Roberts-Borsani2022b, Roberts-Borsani2026}, strongly supporting the UVLF determinations up to these redshifts. The  picture emerging from these surveys is shown in Fig.~\ref{fig:uvlfz9to14}.
We split the measurements into six redshift ranges to minimize the scatter due to the intrinsic evolution in the UVLF.  
In practice, however, most measurements fall within two redshift ranges, $z=9$--11.5 and $z=12.25$--14.5. This is because the redshift windows (and hence the mean redshifts) are set by the JWST-based selection criteria. When the Lyman-break technique is applied to the NIRCam filter set, the sample naturally divides into F115W dropouts ($z\lesssim11.5$) and F150W dropouts ($z\gtrsim12.25$), although photometric redshifts and detailed imaging simulations allow the selection windows to be broadened. 
In Fig.~\ref{fig:uvlfz9to14}, most upper limits arise from the absence of plausible bright candidates in small-area surveys. With the recent completion of relatively wide-area surveys, there are now enough measurements at $M_{\rm UV}\simeq-22$ ($z<12$) and $M_{\rm UV}\simeq-21$ ($z>12$) to provide a first glimpse of the bright end of the UVLF.

The comparison between the various  panels of Fig.~\ref{fig:uvlfz9to14}  shows the evolution of the UVLF across these two redshift ranges,which correspond to cosmic ages of about 450~Myr and 300~Myr, respectively. The evolution is easy to see against our $z=9$ UVLF fit, which is shown in all panels for reference. We quantify this evolution more precisely in the next section, but Fig.~\ref{fig:uvlfz9to14} already shows that it amounts to a factor of $\sim$5 in number density from $z\sim9$ to $z\sim13$--14.

As redshift increases, the luminosity range covered by these analyses shrinks at both the bright and the faint end, owing to the combined effect of fainter apparent magnitudes and the smaller comoving volumes sampled by the surveys. In this context, the data from the first wide-area surveys (e.g., COSMOS,, \citet{Franco2025}, PANORAMIC, \citealt{Williams2025}; BEACON, \citealt{Morishita2025Beacon}) and from the deep surveys in lensing fields \citep{Chemerynska2026,Moreira_Goolsby2026} are essential for sampling the bright and faint ends, respectively, and need to be complemented by future observations. 

However, closer inspection of Fig.~\ref{fig:uvlfz9to14} reveals fluctuations superimposed on the overall evolution. For instance, the points at $z\sim10$ lie somewhat below those at $z\sim11$, and the same inversion appears between $z\sim13$ and $z\sim14$. In particular, the two most recent analyses at $z\sim13$ \citep{McLeod2026,Weibel2026UVLF} lie below the other data at the same redshift, and even below those at $z\sim14$.
We quantify this point more formally in Sect.~\ref{sec:meta-analysis}.

These variations are not surprising, as we discuss below. The measurements are derived from a continuously growing set of imaging surveys that are inhomogeneous in depth and filter coverage. Although all of these studies use broadly similar techniques to derive the UVLF from imaging catalogs, the steps involved are intrinsically complex and uncertain, and their details differ among implementations. Indeed, estimating the UVLF involves much more than ``counting galaxies in a given redshift range and dividing by the cosmological volume.'' It requires some combination of color selection and photometric redshifts to identify robust candidates, together with stringent rejection criteria to remove interlopers and contaminants. Because these greatly outnumber genuine high-$z$ galaxies \citep{Vulcani2017}, all techniques favor purity over completeness, as the earlier HST surveys did, selecting only the most robust candidates in an effort to reject all or most contaminants.

Correcting the samples statistically for the resulting incompleteness requires lengthy simulations, usually at the image level. These simulations are intrinsically difficult and rely on assumptions about the input galaxy populations (both the high-$z$ targets and the lower-redshift interlopers), which makes their results somewhat uncertain. Although most of the studies shown in Fig.~\ref{fig:uvlfz9to14} broadly follow this procedure, their implementations differ in the details, adding to the the Poisson noise and cosmic variance fluctuations expected from the relatively small areas surveyed.

Overall, these findings indicate that, although the evolution of the UVLF at $z>9$ is clearly detected, the scatter among the various determinations is non-negligible and may increase with redshift. This scatter arises not only from statistical fluctuations, including those of physical origin such as cosmic variance, but also from potential systematic biases in the data analysis, which must be taken into account in detailed comparisons between data sets or between data and models.

\subsection{A meta-analysis of the UVLF evolution at \texorpdfstring{$8.5<z<14.5$}{8.5<z<14.5}}\label{sec:fit}
\label{sec:meta-analysis}

To put these considerations on a more quantitative footing, and following the approach of \citet{Bouwens2021} and \citet{Finkelstein2022AGNGAL}, we explore whether the entire evolution in this redshift range can be described by a smoothly evolving function, and what lessons can be drawn from this exercise. To this end, we perform a \enquote{meta-analysis} of all the JWST-based UVLF measurements available in the literature at $8.5<z<15$.

We started with a flexible definition of the fitting formula,  adopting an evolving Schechter function \citep{Schechter1976} where its three parameters $\log(phi)$, $M^*$ and $\alpha$ are allowed to vary as a function of redshift like: 
\begin{equation} 
\log(\phi(z)) = \log(\phi_0) + \log(\phi_1)(z-9) + \log(\phi_2)(z-9)^2;  \qquad M^*(z) = M^*_0 + M^*_1(z-9);  \qquad \alpha(z) = \alpha_0 + \alpha_1(z-9)
\label{Eq:EvolvingSchechter}
\end{equation}

, where $\log(\phi_0)$  $\log(\phi_1)$ and $\log(\phi_2)$ denote the expansion coefficients of $\log\phi$, for a (maximum) total of seven free parameters. In practice, all parameters vary linearly from $z=9$ to $z=14.5$, except $\log\phi^*$, which is allowed to vary quadratically. Given the results described in Sect.~\ref{sec:LF12}, the lack of data at both the bright and the faint end, and the relatively small expected variation of $\alpha$ and $M^*_{\rm UV}$, we deem a linear parameterization sufficient to describe their evolution. The quadratic form for $\log\phi^*$, by contrast, allows us to capture a change in the pace at which galaxies evolve, as suggested by recent analyses \citep{McLeod2026,Weibel2026UVLF}.

All the free parameters of $\log\phi^*$, $M^*_{\rm UV}$, and $\alpha$ are obtained through a standard minimization procedure against the data available in the fitted redshift range. For this purpose, we use the data from \citet{\UVLFfit}.

The Schechter form that we adopt is somewhat arbitrary. For instance, there are indications that the UVLF at somewhat lower redshift is better described by a DPL \citep[e.g.,][]{Bowler2020, Finkelstein2022CANDELS}.  However, given the weak constraints at the brightest magnitudes, we find no reason to introduce a DPL in the present analysis, and we therefore adopt the Schechter function, that has less free parameters. Indeed, other studies at these redshifts use both simultaneously, given the lack of strong evidence in favor of either \citep{Harikane2025, Shuntov25, Whitler2025,McLeod2026}.

More importantly, we emphasize that this procedure is necessarily coarse and does not constitute a rigorous analysis of the UVLF evolution, for several reasons. First, many points are not independent, because they derive from different reanalyses of the same data sets. Second, many technical aspects of the procedures used to estimate the UVLF differ among authors, introducing systematic uncertainties that should not be treated as statistical ones. Despite these concerns, we take an agnostic approach, assuming that all published analyses are equally valid and that together they bracket the overall uncertainties currently affecting this measurement, and weighting each result by its reported errors. With these caveats, and others detailed below, we find that this exercise provides a useful view of the global evolution of the UVLF, and we have performed two different analyses with this method, that are described below. Details of the fitting procedure and its results are given in the Appendix.

First, we obtained a global \enquote{evolutionary} fit using all the UVLF data over the entire range from $z=9$ to $z=14.5$. Unsurprisingly, we found that the form adopted in Eq.~\ref{Eq:EvolvingSchechter} has too many free parameters to be constrained by the data, given the poor sampling of the faint and bright ends discussed in Sect.~\ref{sec:LF12}. After experimenting with fixing several combinations of these parameters, we adopted a sequential procedure in which we progressively fixed to zero the evolutionary terms found to be statistically consistent (within $2\sigma$) with zero. With this procedure, a smooth and statistically acceptable fit is obtained with only four free parameters: a constant $\alpha\simeq-2.3$, a constant $M^*_{\rm UV}\simeq-21.7$, and a linearly evolving $\log\phi^*(z)\simeq-4.8-0.17\,(z-9)$. All other terms are consistent with zero. We note that the values of these four parameters are remarkably stable when we add further free parameters.

We adopt this solution as our \enquote{Fiducial} model and use it in the following.

 The results  are presented in Fig.~\ref{fig:uvlfz9to14}, while the best-fit parameters and other technical details are given in the Appendix and in its Tab.~\ref{tab:schechter_params}. 

\begin{wrapfigure}{r}{0.5\textwidth}
\centering
\includegraphics[width=0.95\linewidth,keepaspectratio]{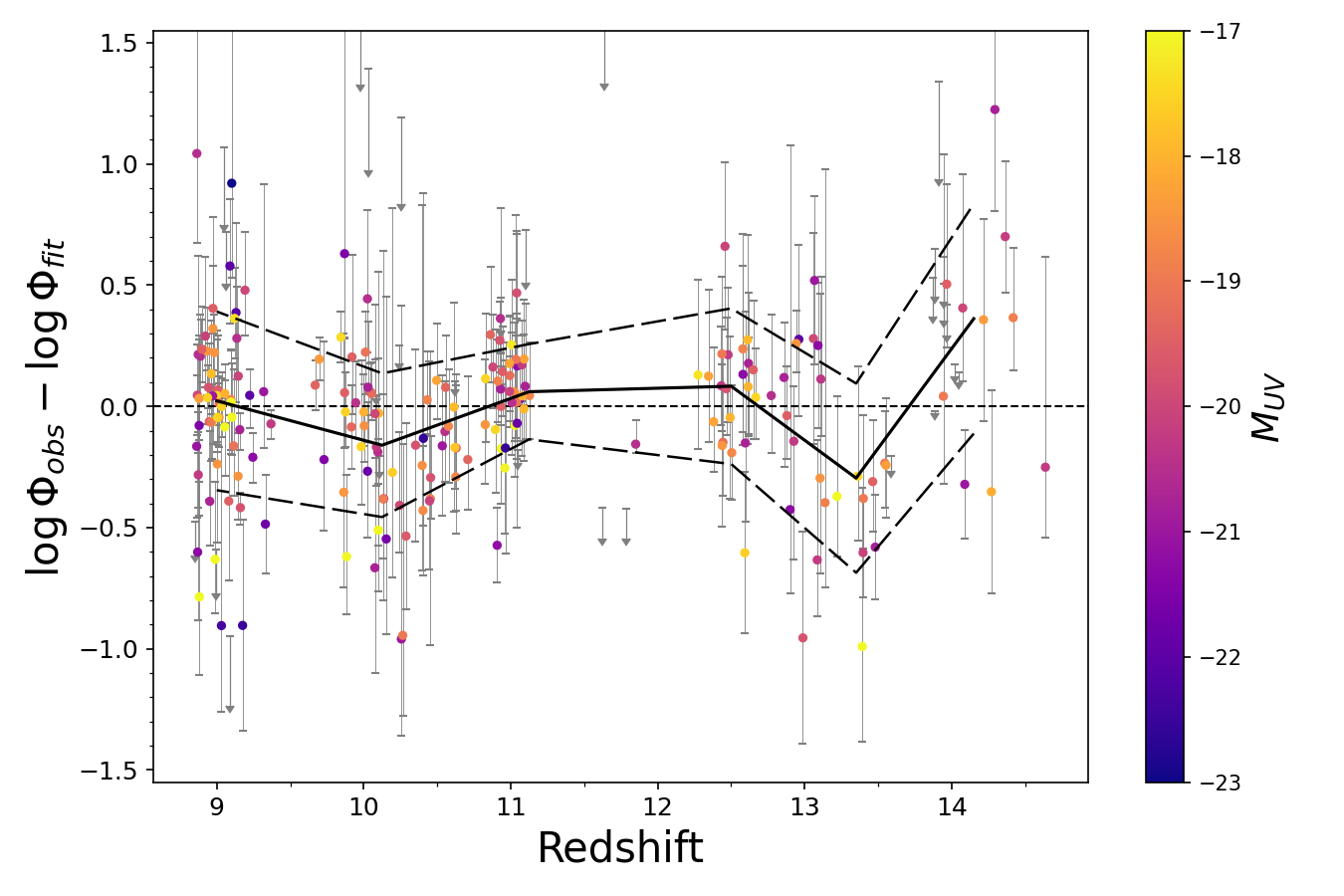}
\caption{Residuals of the observed UVLF measurements with respect to the \enquote{Fiducial} evolving Schechter fit. The lines show the median and the $1\sigma$ scatter of the residuals. For clarity, points are randomly offset in redshift by up to $\pm0.15$, and are color-coded by absolute UV magnitude $M_{\rm UV}$, as indicated by the color bar. Upper limits have not been used to calculate the scatter.}
\label{fig:uvlfresiduals}
\end{wrapfigure}

The main findings for the four-parameter \enquote{Fiducial} model are:
\begin{itemize}
\item The four-parameter Schechter model provides a good description of all the data, with $\chi^2_{\rm red}\simeq1.35$. Reducing the number of free parameters from seven to four does not reduce the goodness of the fit.  For about 200 degrees of freedom, this excess over unity corresponds formally to a $3.3\sigma$ deviation from the value expected if the reported errors were purely statistical. This indicates additional sources of error, consistent with the inter-study systematics discussed above and also further discussed below (Fig.~\ref{fig:uvlfresiduals}).
    \item The slope $\alpha$ is  steep ($\alpha_0\sim-2.31^{+0.05}_{-0.04}$), in line with the results of \cite{Moreira_Goolsby2026} and somewhat steeper than \cite{Chemerynska2026}. We note that the sampling of the UVLF slope is limited especially at $z>12$, which may explain why there is no evidence for any evolution in $\alpha$.
    \item The faint-end slope is steep ($\alpha_0=-2.31^{+0.05}_{-0.04}$), in line with the results of \citet{Moreira_Goolsby2026} and somewhat steeper than that of \citet{Chemerynska2026}. The faint end of the UVLF is poorly sampled, especially at $z>12$, which may explain why we find no evidence for evolution in $\alpha$.
    \item The dominant term is the evolution of the normalization, $\phi_1\equiv\Delta\log\phi^*/\Delta z=-0.17\pm0.01$, corresponding to a decline in number density by a factor of $\simeq1.5$ per unit redshift. This trend is robust: we verified that it cannot be reproduced by evolving the other parameters instead, despite the correlations among them.
    \item The linear evolution of $\log\phi^*$ provides a good fit to the data, with no evidence that the rate of density evolution changes anywhere in the range $z=9$--14.5. Not only do the fit residuals show no sign of such a change (see below), but when we add a quadratic term $\phi_2$ to the density evolution, its best-fit value is consistent with zero. There is, however, an important caveat regarding the results from most recent surveys that is discussed below.
\end{itemize}

We also performed a complementary analysis, fitting the same UVLF data separately in six redshift bins (centered at $z=9$, 10, 11, 12.25, 13.1, and 14), chosen where most of the data lie (see Fig.~\ref{fig:uvlfresiduals}). In this case, we fit only the three Schechter parameters $\log\phi^*$, $M^*_{\rm UV}$, and $\alpha$ in each bin. The resulting values are also reported in Table~\ref{tab:schechter_params}. Unlike the evolving fit, these fits show a redshift dependence in both $\phi^*$ and $M^*_{\rm UV}$: the milder evolution of $\phi^*$ (from $\log\phi^*\simeq-4.9$ at $z=9$ to $\log\phi^*\simeq-5.15$ at $z=14$) is compensated by $M^*_{\rm UV}$ becoming fainter (from $M^*_{\rm UV}\simeq-21.75$ at $z=9$ to $M^*_{\rm UV}\simeq-21$ at $z=13$--14), albeit with statistical fluctuations between bins. Remarkably, $\alpha\simeq-2.3$ in every bin, confirming the value from the four-parameter fit. The differences in the evolution of $\phi^*$ and $M^*_{\rm UV}$ between the two approaches are well within the uncertainties and the degeneracies among the Schechter parameters (see Fig.~\ref{fig:uvlf_cornerplot}), and do not undermine the validity of the exercise.
Clearly, these differences would change the physical interpretation if the best-fit values were used to distinguish between evolutionary modes of the UVLF (e.g., luminosity versus density evolution) or to identify a differential evolution of bright and faint galaxies (which would manifest as a decline of the characteristic luminosity). However, as Fig.~\ref{fig:uvlfz9to14} shows, the two sets of UVLFs are nearly indistinguishable over the magnitude range covered by the data. This confirms that the differences are compatible with the known degeneracies among the Schechter parameters, and suggests that they result from the inadequate coverage of the UVLF, especially at the bright end, which prevents us from deriving physically robust parameters. For this reason, we use these fits only as a smooth representation of the data. 
In the following, we adopt the \enquote{Fiducial} evolving fit, which uses all the data at once and can be evaluated continuously at any redshift with only four parameters.

Overall, these results confirm the findings discussed in the previous section. The data point to a steady decline in the number density of high-redshift galaxies, that is of about $0.85$dex \enquote{Fiducial} evolving fit between $z=9$ and $z=14$. They also expose the limitations of current surveys. The bright end of the UVLF remains poorly constrained, particularly at $z>12$, and wide-area surveys are required for a more robust analysis of the evolution of the brightest galaxies. We note, however, that the very detection of three spectroscopically confirmed galaxies at $z\geq14$ \citep{Carniani2024a, Naidu2025} sets a robust lower limit on the UVLF at $z\sim14$, at least for the area surveyed so far. This result needs to be validated by future surveys.

In this context, we note a non-negligible tension between the picture derived from this \enquote{meta-analysis} and two recent analyses at $z\sim13$ \citep{McLeod2026,Weibel2026UVLF}. These two studies are based on the largest-area data sets to date at these redshifts, and both yield UVLFs that are on average lower than our \enquote{Fiducial} fit (even though they are included in the \enquote{meta-analysis}) and than the other data at these redshifts. These analyses may well provide a more accurate determination of the UVLF than previous ones, as they benefit from the largest areas. However, methodological differences among studies may also play a role, as we discuss in more detail below. For instance, the number density derived from the PANORAMIC data alone \citep{Weibel2026UVLF} lies above the average over all fields, most of which are included in our analysis. This suggests that the difference is not due to the wide-area PANORAMIC data, but more generally to the overall procedure of candidate selection and UVLF estimation. The densities measured by these two analyses are also below those of most other studies at $z\sim14$; the other wide-area survey \citep{Kreilgaard2026} yields only upper limits, which are consistent with the other data at $z=14$. Addressing this discrepancy is a priority for future analyses.

Similarly, the faint end is poorly constrained, especially at $z>12$. The slope at very faint limits and the potential existence of a turnover at faint magnitudes has a number of important consequences \citep[e.g.,][and references therein]{Castellano2016c, Sipple2024ApJ} and is only starting to be explored by the deepest data available today \citep{Chemerynska2026, Moreira_Goolsby2026}. More deep surveys of lensing fields, backed by spectroscopic follow-up, are needed to validate this important result. Programs such as GLIMPSE \citep{Asada2026spec}, SPURS (GO~9214, PIs Mason and Stark; \citealt{Chen2026SPURS}), and the forthcoming DAJE (GO~9645, PIs Treu, Fontana, and Roberts-Borsani) will mark a crucial step forward in this direction.

The fit residuals provide further insight into the evolution of the data. Figure~\ref{fig:uvlfresiduals} shows the residuals between the observed data used in the fit and the best-fit function as a function of redshift, together with their median and rms scatter, which quantify the agreement among the various studies. The median residuals show no strong trend with redshift, supporting the quality of the fit. However, the residuals have an rms scatter of $\Delta\log\phi\simeq0.1$--0.2 at all redshifts, which increases beyond $z\sim13$. They also show systematic offsets between the determinations at $z\simeq13$ and $z\simeq14$, which lie systematically below and above the fit, respectively. These deviations can plausibly be explained by a combination of smaller samples at these redshifts, incomplete filter coverage in some wide-area surveys, and differences in the technical details of the UVLF estimation.

These residuals indicate that individual studies are still affected by a level of systematic uncertainty comparable to the observed density evolution per unit redshift ($\sigma(\log\Phi)\simeq\Delta\log \phi/dz\simeq 0.2$). Caution is therefore needed in detailed comparisons between studies, and between individual studies and theoretical models.


\section{The implications for galaxy formation models}\label{sec:theory}

The search for galaxies at the earliest epochs is not (only) an exploration of the Universe back to its beginnings: it is an essential part of a long-term effort to build a global theory of the birth and evolution of galaxies \citep[see, e.g.,][for a review]{Dayal2018}. The key tool for this goal is the development of \textit{ab initio} theoretical models, i.e., models that attempt to describe self-consistently the growth of early galaxies from the primordial fluctuations of a well-defined cosmological model \citep[see][and references therein, for recent reviews of the underlying physical implementation]{Dayal2018,Somerville2026}. A remarkable feature of these models is that the galaxy population closely traces the population of dark matter (DM) halos, since the mean number of galaxies per halo of a given mass is close to unity at high redshift. In this sense, the growth of galaxies in the first 500~Myr is driven by the growth of the underlying DM halos. While this growth, as described by the halo mass function, is usually considered a solid theoretical prediction, numerical uncertainties remain in calculating it at very early epochs \citep[$z\gtrsim15$--20;][]{Somerville2025}, adding to the uncertainties and discrepancies among theoretical models.
That said, the ambition of these models is to describe the complex set of physical mechanisms that govern the formation of the first galaxies, including the full cycle of gas cooling and reheating, star formation from the first generation (see the recent review by \citealt{Venditti2026}) to more enriched populations, and radiative and mechanical feedback, all within the framework of the complex dynamical interactions occurring within and among galaxies in the dense high-redshift Universe \citep[e.g.,][]{Jaacks2019,katz2025}. These processes produce the first radiative sources, so the luminosity of the first galaxies also depends on the properties of these sources, such as the initial mass function (IMF), the fraction and impact of binaries and Wolf--Rayet stars, and the presence and radiative properties of SMBHs, including those potentially formed through non-standard, non-stellar channels. All these aspects are, at least in principle, included in these models (although not all of them in all models), either through self-consistent prescriptions or as input parameters.

This section presents an updated comparison between the data and the models developed after the first JWST results. In addition to these self-consistent models, other studies have used simpler scaling relations to investigate the effect of individual factors, such as a change in the mass-to-light ratio due to stellar age \citep{mason2023,Donnan2024} or a mass-dependent scatter in UV luminosity due to bursty star formation \citep{Gelli2024, nikopoulos2026}. These analyses are important for isolating the role of specific processes, but are not discussed here.

\subsection{A new generation of models}
\label{sec:theory-model}

\begin{figure*}[ht]
\centering
\includegraphics[width=0.95\linewidth,keepaspectratio]{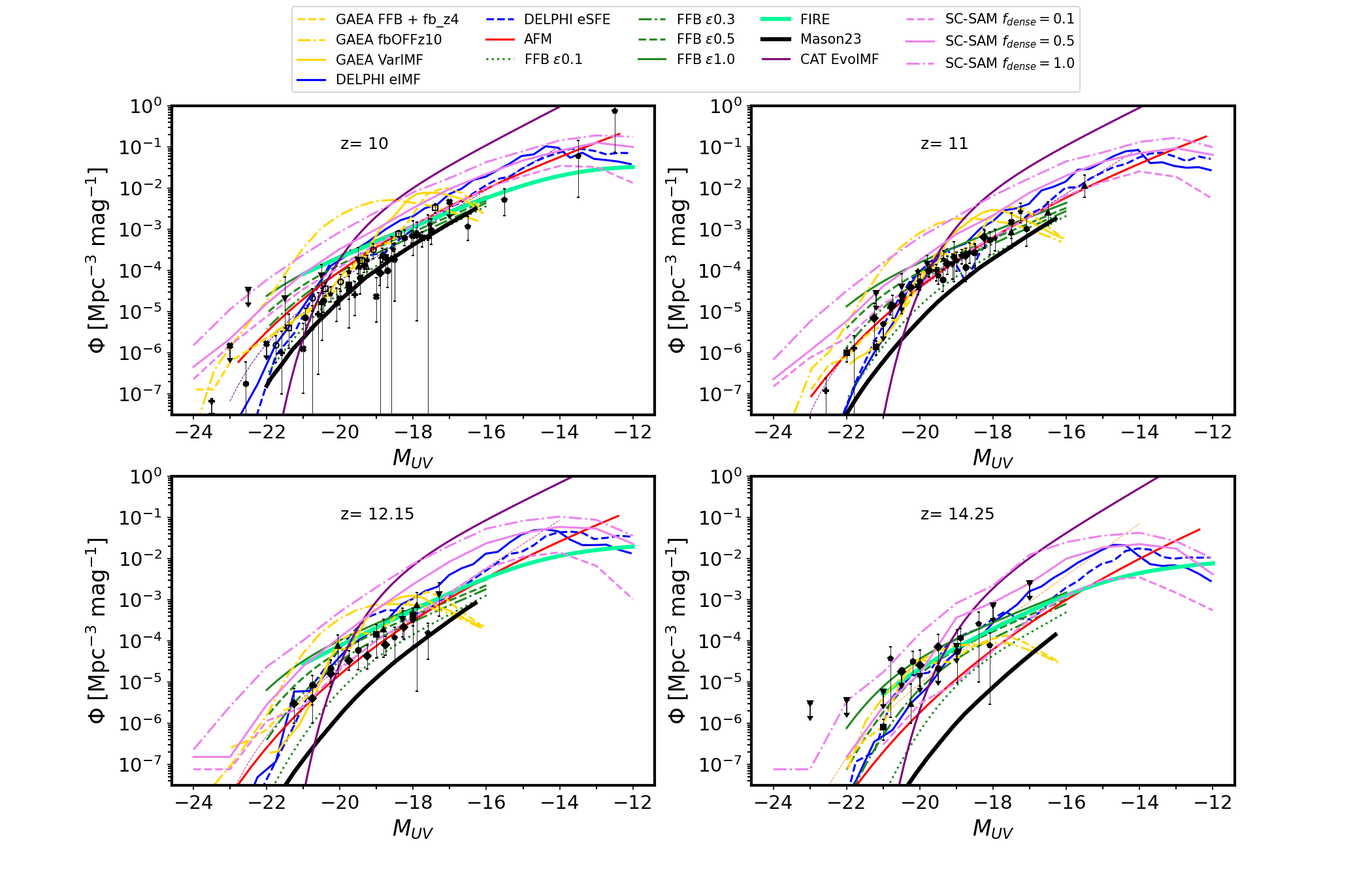}
\caption{Comparison between the observed measurements of the UVLF and a suite of post-JWST theoretical models. Data are the same used in Fig\ref{fig:uvlfz09to10}. Theoretical models (GAEA FFB$+$fb\_z4 and GAEA fbOFFz10: \cite{Cantarella2025}; GAEA VarIMF: \cite{Fontanot2026}; DELPHI:  Mauerhofer2025, AFM: \cite{Ferrara2026UVRED}; FFB: \cite{Li2024}; FIRE: \cite{Samuel2026}, CAT: \cite{Trinca2024}; SC-SAM: \cite{Somerville2025}) are shown with lines with different colors and styles as shown in the legend. We also show for comparison the pre-JWST model of \cite{mason2023}. Further details are given in the text. }
\label{fig:models}
\end{figure*} 

As soon as the first galaxies beyond $z=10$ were detected, it quickly became clear that these findings challenged most existing theoretical models. Given that most models reproduce the observed galaxy demographics at $z\sim5$--10, the expectation was that they would easily yield a progenitor population consistent with the higher-redshift observations. This expectation was based on the perhaps naive assumption that galaxy physics becomes simpler at earlier stages. The most comprehensive comparison between early JWST observations and pre-JWST models is presented in Fig.~8 of \citet{Finkelstein2024}, to which we refer the reader. As shown there, essentially all pre-launch models underpredict the number of galaxies brighter than $m_{\rm AB}\sim28.5$ at $z\geq11$, by an amount that increases with redshift.

This discrepancy has motivated the revision of many of these models, to include or modify physical processes that may have played a significant role in the growth of early galaxies. These revisions did not rely on \textit{ad hoc} recipes or additional free parameters tuned to match the data. Instead, they identified several processes that may have been effective at $z>10$ and that become less important, or negligible, at lower $z$. These mechanisms result from the peculiar conditions of the very high-redshift Universe, where higher matter densities, shorter dynamical timescales, and pristine or very metal-poor gas and stellar populations may easily have produced physical conditions not typically seen at lower $z$.

The astrophysical solutions that have been proposed so far include:\\
$\bullet$ A decrease in the dust attenuation of early sources, allowing a larger fraction of the intrinsic UV light to reach the observer. The decrease in dust attenuation could either be due to radiation-driven outflows of dust \citep{ferrara2022, Ferrara2024, nakazato2025}, dust being dispersed over a larger fraction of the virial radius \citep{mauerhofer2023, nikopoulos2026} or being spatially segregated from star formation sites \citep{Ziparo2023}. An evolution of dust properties may also have a role in this context \citep{Burgarella2026,Narayanan2026O}.\\
$\bullet$ An increase in the star formation rate efficiency, modulated either by density \citep{Somerville2025}, ineffective feedback \citep{Dekel2023, renzini2023, Li2024}, or redshift as implied by hydrodynamic simulations \citep{Mauerhofer2025}.\\
$\bullet$ A bias in flux-limited observations leading to the preferential detection of galaxies with extreme star-bursts or stochastic star formation \citep{mason2023, mirocha2023, Munoz2023, sun2023,Yung2024, Gelli2024,nikopoulos2026, Munoz2026}.\\
$\bullet$ A stellar Initial Mass Function  that becomes increasingly top-heavy with increasing redshift, resulting in high light-to-mass ratios \citep{Trinca2024, cueto2024, Yung2024, Menon2024, jeong2025, Ventura2025}.\\
$\bullet$ A larger fraction of rapidly rotating stars in the first galaxies, thus boosting the number of UV photons per stellar baryon \citep{Liu_rot2025}. This boost is achieved because the rapid rotation prevents the stars from evolving off the main sequence to the red part of the Hertzsprung-Russel Diagram. Such rapid rotation in low-metallicity stellar populations is predicted in terms of star formation pathways \citep{Stacy_rot2011,Stacy_rot2013}, and of significantly reduced angular-momentum loss in the absence of radiatively-driven winds \citep[e.g.,][]{Kudritzki2002,Vink_wind2022}.\\
$\bullet$ A contribution from SMBH - of different origins - that may be lurking in the galaxy and are not easily identified as a bright, broad-line AGN \citep{pacucci2022,Hegde2024}. In turn, this is potentially linked  to the claimed  large abundance of AGN at lower redshifts and the overmassive BH results at  lower redshifts. 


Although these models can predict many quantities to be compared with the data, the UVLF remains the most direct and fundamental test of this class of models, and in this review we limit our analysis to it. In particular, we consider the predictions of the following recent post-JWST models:\\
$\bullet$ \textbf{AFM} \citep{Ferrara2026UVRED}. Strictly speaking, this is not truly an \enquote{ab initio} model, but rather a physically motivated analytic or semi-numerical framework in which the galaxy--halo relation, star formation, and feedback are described by a small set of prescriptions calibrated at lower redshift. The model explains the unexpectedly high abundance of UV-bright galaxies at $z>10$ without invoking extreme star-formation efficiencies. When the rapidly increasing specific star-formation rate exceeds a dust-modified Eddington limit, radiation pressure drives powerful outflows that clear dust from the star-forming regions, strongly reducing UV attenuation and producing a brighter, attenuation-free phase. This model should not, however, be considered \enquote{dust-free}: the coupling between dust and radiation controls the transition between the obscured and unobscured phases. The model therefore predicts an evolution from an initially dust-obscured \enquote{Red Monster} phase to an exposed, UV-luminous \enquote{Blue Monster} phase, and hence the possible coexistence of very blue and heavily obscured systems at similar redshifts. The kinematic and spatial signatures of these outflows provide direct observational tests of this scenario.\\
$\bullet$ \textbf{CAT} \citep{Trinca2024}. This is a semi-analytical model developed to follow the co-evolution of the first galaxies and their nuclear BHs from the earliest cosmic times. Its basic prescriptions include a specific treatment of the evolution and impact of Pop~III stars, described in detail by \citet{Trinca2022}. The version adopted here \citep{Trinca2024} includes a gradual evolution of the IMF, following recent 3D hydrodynamical simulations of low-metallicity star-forming regions \citep{Chon2021} that show a smooth transition in the IMF up to a metallicity of $Z\sim10^{-2}\,Z_\odot$. As a result, high-redshift, low-metallicity young galaxies may host a larger fraction of massive stars than predicted by a standard Salpeter-like IMF.\\
$\bullet$ {\bf DELPHI} \citep{Mauerhofer2025, dayal2026b}. The DELPHI SAM has been used for the study of early galaxies since its development in 2014. This model uses analytical merger trees to study the assembly of halos ranging between $10^{8-14}M_\odot$ at $z \sim 4.5$, whose assembly is followed out to $z \sim 40$. It includes all of the key baryonic physical processes of gas and stellar mass assembly through mergers and accretion, star formation and its associated feedback, and multiple black hole seeding pathways. It also includes state-of-the-art dust and metal enrichment calculations \citep{dayal2022}, motivated by results from ALMA observations at $z 
\sim 5-7$. The latest version incorporates ISM physics, namely the cold gas fractions and star formation efficiencies in halos ``learnt" from the SPHINX hydrodynamic simulations \citep{rosdahl2022}. The authors explore two approaches for solving the over-abundance of bright systems at high-z: (i) allowing an increase in the star formation efficiency coupled with lower dust attenuations with increasing $z$; and (ii) allowing an IMF that becomes increasingly top heavy with increasing $z$, based on hydrodynamic simulations of early star formation \citep{chon2022}. Crucially, the authors highlight how observables such as the ionizing photon production rate per unit UV light, and the UV spectral slopes, can be used to distinguish between these two classes of models.    \\
$\bullet$ \textbf{FFB} \citep{Li2024}. The key feature of this model, originally proposed by \citet{Dekel2023}, is the inclusion of feedback-free starbursts (FFBs). These occur whenever the gas density exceeds a critical threshold of $\sim3\times10^{3}\,{\rm cm^{-3}}$, which happens with increasing probability at high redshift. At these densities, the free-fall time is $<1$\,Myr, shorter than the delay between a starburst and the onset of effective feedback from supernovae, as well as from stellar winds when the metallicity is below $\sim0.2\,Z_\odot$, a condition also typical of high redshift. If the starbursts occur in clouds of size $\gtrsim10$\,pc, the surface density is high enough to prevent radiative feedback from disrupting the FFB process.\\
$\bullet$ \textbf{FIRE} \citep{Samuel2026}. This model is based on a suite of cosmological hydrodynamic simulations of early galaxy formation ($z\geq6$) from the Feedback In Realistic Environments (FIRE) project. In particular, this version uses the FIRE-3 model \citep{Hopkins2023} and combines the large volume and statistics of the BonFIRE simulation ($L\sim40$\,cMpc, $m_{\rm b}\sim5\times10^4\,M_\odot$) with the higher resolution of the CampFIRE simulation ($L\sim5$\,cMpc, $m_{\rm b}\sim0.8$--$6\times10^3\,M_\odot$) to robustly predict galaxy properties over a wide dynamic range ($M_*\sim10^4$--$10^{10}\,M_\odot$). Notably, the model also includes a simplified treatment of the IMF of Pop~III star particles, assuming that the IMF becomes top-heavy below a metallicity of $Z=10^{-5}Z_\odot$.\\
$\bullet$ \textbf{GAEA} \citep{Cantarella2025, Fontanot2026}. These models are the latest versions of the semi-analytic model described in detail by \citet{DeLucia2024}. Compared with the original model, they include more refined treatments of chemical enrichment, stellar feedback, the partitioning of gas into its atomic and molecular components, and cold gas accretion onto and outflows from supermassive BHs. For \citet{Cantarella2025}, we consider the further revisions introduced to better match the high-$z$ data, namely: \textit{a)} a model \enquote{FFB + fb\_z4}, which implements a version of the FFB model of \citet{Dekel2023} and \citet{Li2024} combined with weaker stellar feedback at $z>4$ (purple line in \citealt{Cantarella2025}); and \textit{b)} a model \enquote{fbOFFz10}, in which gas reheating and energy injection are completely switched off at $z>10$. \citet{Fontanot2026} adopt the same version of the GAEA model, introducing a physically motivated prescription for a varying stellar IMF. This prescription is inspired by several lines of evidence, including a number of results for high-redshift galaxies \citep{Katz2022,Cameron2024,Cullen2025,Curti2025}.\\
$\bullet$ \textbf{SC-SAM} \citep{Somerville2025}. The Santa Cruz semi-analytic model has been used for more than 25 years to predict many aspects of galaxy evolution. Its core physical prescriptions are described in detail by \citet{Somerville2008,Somerville2015}, and its latest parameter calibration by \citet{Yung2024}. The version adopted here \citep{Somerville2025}, dubbed Density-Modulated Star-Formation Efficiency (DMSFE), includes a star-formation efficiency that increases with gas surface density. This scaling stems from the assumption that at least some of the stars in high-redshift galaxies form in clouds or bound clusters that are much denser than those in normal nearby galaxies, an assumption supported by observations of extremely compact star-forming clumps and by high-resolution hydrodynamical simulations. The process is regulated by a free parameter $f_{\rm dense}$ (hereafter $f_{\rm d}$), which represents the fraction of the ISM in bound protoclusters or dense star-forming clouds.\\

Figures~\ref{fig:models} and \ref{fig:deltamodel} compare several of these updated models with the latest data over the whole redshift range solidly covered by JWST observations, $z=10$--14.5. The two figures tell essentially the same story, but their different perspectives offer complementary insights into the behavior of the models. 
Before we dive into the analysis of the most recent models, we note that Fig.~\ref{fig:models} also shows an example of a pre-JWST model, namely the analytic model of \citet{mason2023}. This model uses the halo formation timescales derived from extended Press--Schechter theory as a proxy for the star-formation rate (SFR), combined with a reasonable assumption about the resulting star-formation efficiency, to provide a simple expectation for the observed number density of galaxies. As shown in Fig.~\ref{fig:models}, this simple prediction underpredicts the observed numbers, illustrating the need for new physical prescriptions.

Figure~\ref{fig:models} shows the standard comparison between these models and the UVLFs from recent surveys (the data are those shown in Fig.~\ref{fig:uvlfz09to10}). We often limit the comparison to a subset of the models analyzed by these authors, selecting those that best fit the available data. In many cases, the authors explore a wider range of parameters, some of which are in strong tension with the data. We omit these cases for clarity, and refer the reader to the original papers for a full analysis. Because the UVLF evolves rather quickly in this range, and data and models are typically available at specific redshifts, we have grouped most of the existing observations into four relatively narrow redshift ranges, chosen to contain most of the data (see Fig.~\ref{fig:uvlfresiduals}). Even so, there may be a slight offset between the mean redshift of the data and that at which the models are computed, as detailed in the caption of Fig.~\ref{fig:models}.

In Fig.~\ref{fig:deltamodel}, we instead select two specific rest-frame absolute magnitudes, $M_{\rm UV}=-20.5$ and $-18$, representative of luminosities well sampled by existing data, and plot the logarithmic difference between the model predictions and the data as a function of redshift. To provide a smooth representation, we describe the data with the evolving Schechter fit described above (Eq.~\ref{Eq:EvolvingSchechter}). We also show the observed data points at approximately the same $M_{\rm UV}$. For consistency with the models, we plot the difference between each data point and the Schechter fit evaluated at the exact $M_{\rm UV}$ and redshift of that point. This comparison avoids the redshift uncertainty present in Fig.~\ref{fig:models}. Figure~\ref{fig:deltamodel} also shows a $\pm0.2$\,dex band,  which  is the typical amount of  statistical and systematic noise  in the Schechter fit (see discussion in Sect.\ref{sec:fit}).

Several conclusions can be drawn from this comparison:
\begin{itemize} 
\item Including new processes significantly increases the predicted number of galaxies over the entire redshift range considered, to the point that the observed densities lie well within the range of predicted values. \textit{In summary, the observed luminosities and number densities of galaxies up to $z\sim14.5$ can easily be explained by a variety of physical models}.
\item  We would even venture to say, looking in particular at Fig.~\ref{fig:deltamodel}, that the attempt to reproduce the observed number densities has been almost too successful, as most of the models shown there \textit{overestimate} the observed densities over a wide range of redshifts and luminosities. This is particularly true at the bright end of the UVLF, as is evident from the lower panel of Fig.~\ref{fig:deltamodel}.
\item The current scatter among theoretical predictions is significantly larger than the uncertainties in the observed measurements, even when the statistical and systematic effects discussed above are included.
\end{itemize}

\subsection{A wealth of new physical processes}
\label{sec:theory-process}

\begin{figure*}
\centering
\includegraphics[trim={1cm 0.5cm 2cm 0.cm},clip,width=0.33\linewidth,keepaspectratio]{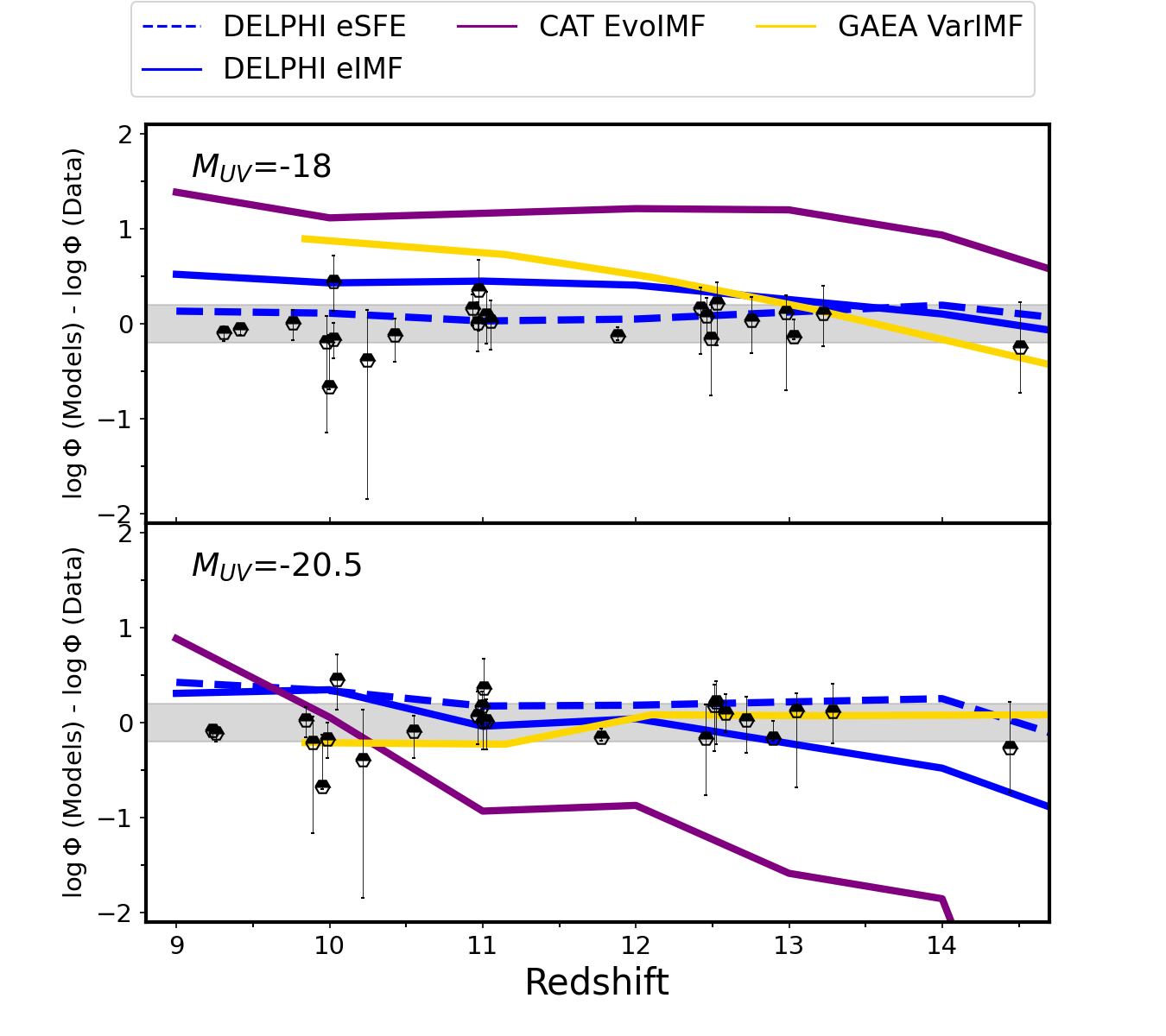}
\includegraphics[trim={1cm 0.5cm 2cm 0.cm},clip,width=0.33\linewidth,keepaspectratio]{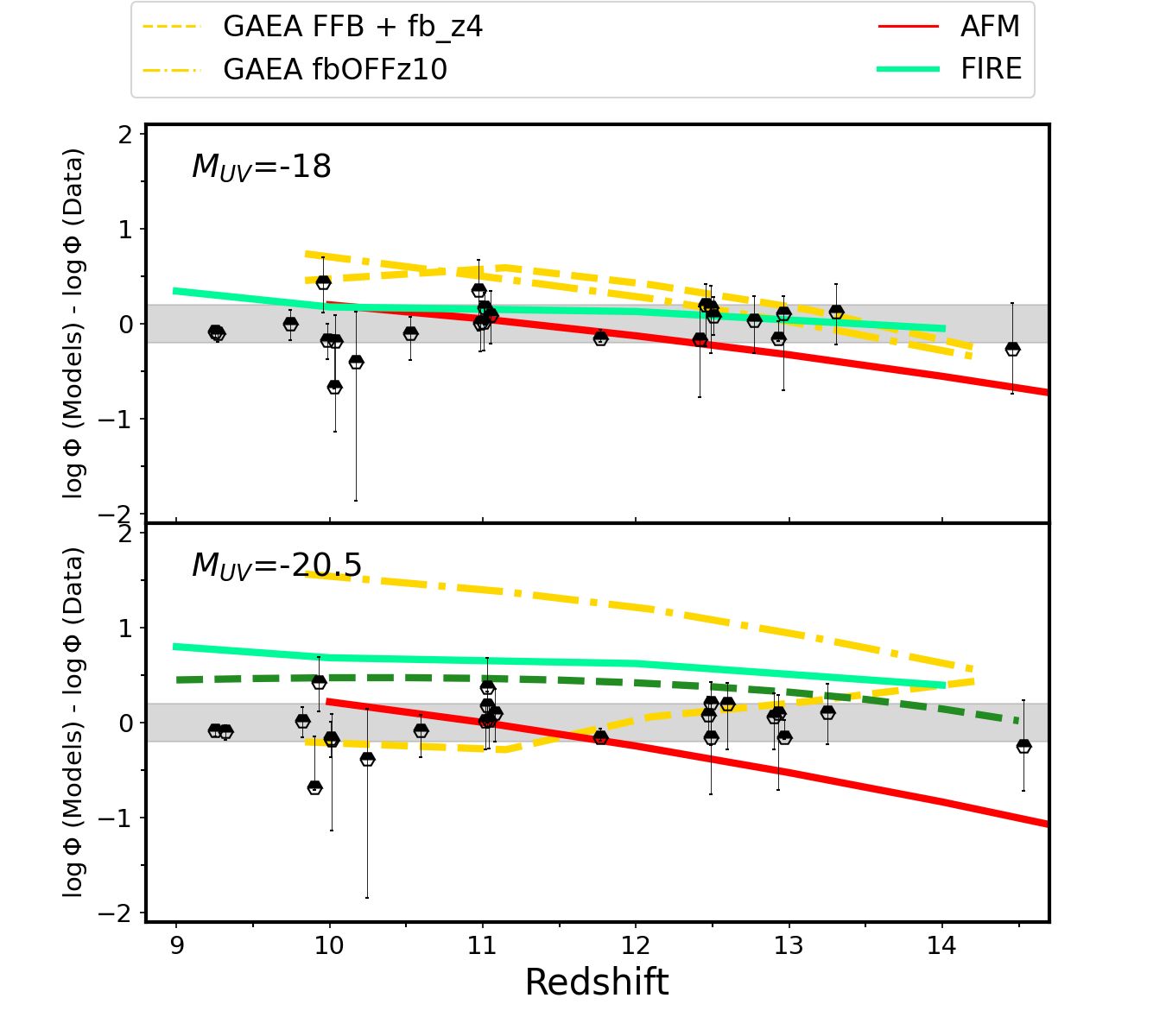}
\includegraphics[trim={1cm 0.5cm 2cm 0.cm},clip,width=0.33\linewidth,keepaspectratio]{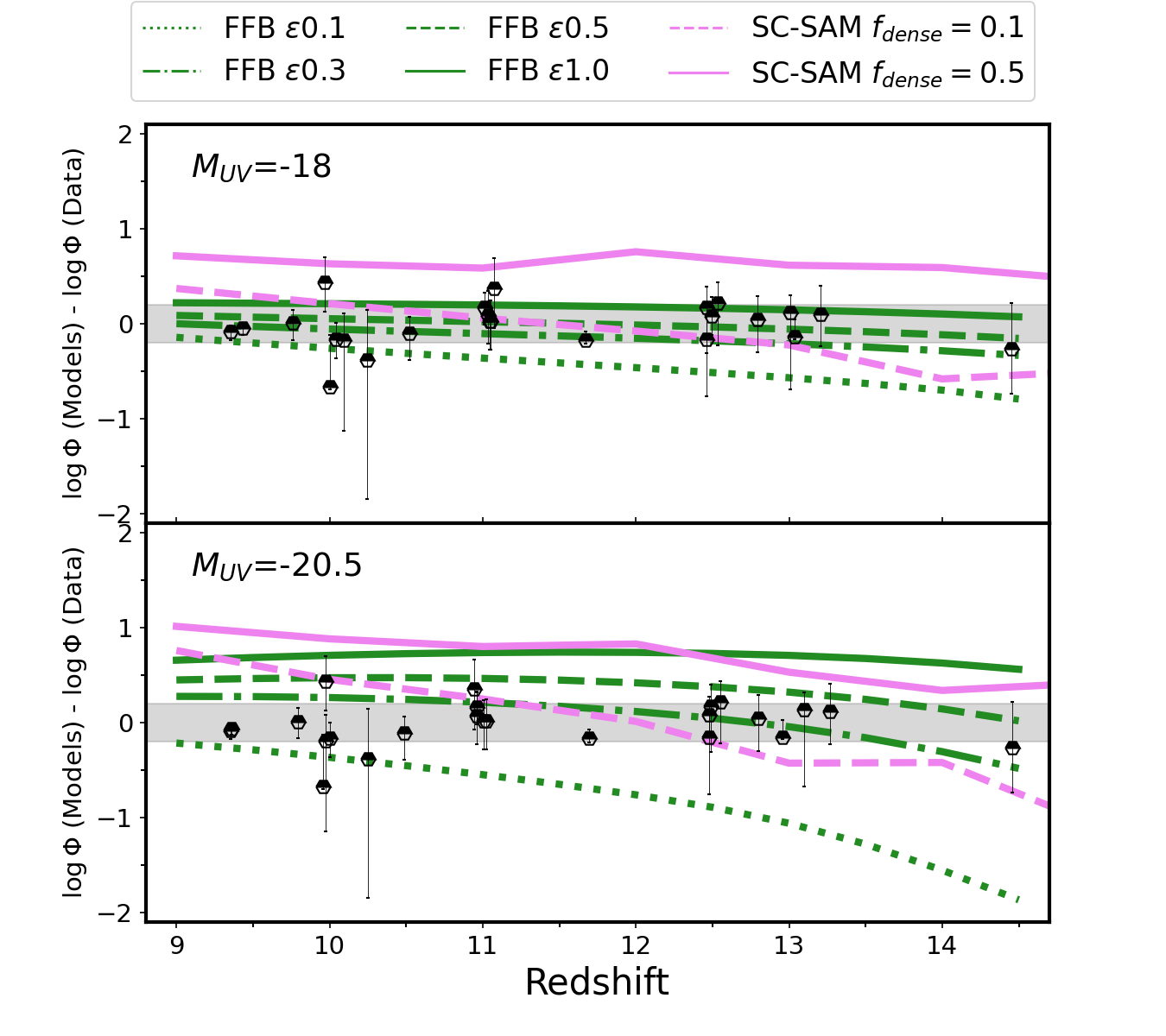}
\caption{Comparison between data and models, with symbols and lines as in Fig.~\ref{fig:models}. The figure shows the logarithmic difference (i.e., the logarithm of the ratio) between the galaxy densities predicted by the models and those observed, as a function of redshift, at two absolute magnitudes: $M_{\rm UV}=-18$ (upper panels) and $M_{\rm UV}=-20.5$ (lower panels). These are broadly representative of faint and bright galaxies, respectively, and data are available at both magnitudes across the entire redshift range considered here. For clarity, the models are split into three panels. The left panel shows models in which the IMF varies with redshift (CAT, \citealt{Trinca2024}; DELPHI, \citealt{Mauerhofer2025}; GAEA VarIMF, \citealt{Fontanot2026}). The central panel shows models that implement specific revisions to the feedback efficiency (AFM, \citealt{Ferrara2026UVRED}; GAEA FFB$+$fb\_z4 and GAEA fbOFFz10, \citealt{Cantarella2025}), together with the FIRE hydrodynamical simulations \citep{Samuel2026}. The right panel shows models in which the star-formation efficiency is increased by various mechanisms (FFB, \citealt{Li2024}; SC-SAM, \citealt{Somerville2025}). See the text for details.}
\label{fig:deltamodel}
\end{figure*} 

It is also tempting to compare cases in which different theoretical approaches, which may differ in several implementation details, introduce similar physical effects. To facilitate this, we split the model comparison in Fig.~\ref{fig:deltamodel} into three panels. For instance, the left panel shows a set of models (CAT, DELPHI-eIMF, and GAEA-VarIMF; \citealt{Trinca2024, Mauerhofer2025, Fontanot2026}, respectively) that explore the possibility of an \textit{evolving IMF}, which becomes progressively more top-heavy at high redshift. These models are motivated by numerical simulations of the birth of star clusters at high redshift, which show that high gas densities and low metallicities, typical conditions for JWST sources, favor top-heavy IMFs \citep{Chon2021, Chon24}. A top-heavy IMF naturally increases the light emitted per unit stellar mass formed, boosting the luminosity of galaxies at a given SFR. However, some of these models also show that this effect is partly (or largely) offset by stronger supernova (SN) feedback, which reduces the overall star formation \citep[e.g.,][]{jeong2025}.

The left panel of Fig.~\ref{fig:deltamodel} shows that the two most recent models (DELPHI-eIMF, \citealt{Mauerhofer2025}; GAEA-VarIMF, \citealt{Fontanot2026}) give similar predictions over the entire range of redshifts and luminosities explored here, and are among the models that best reproduce the data, making this effect a promising candidate explanation. This is in line with several observational hints in this direction, including enhanced production rates of ionizing photons relative to the UV luminosity \citep{simmonds2024}, extremely blue UV slopes suggestive of young, dust-free populations \citep{Cullen2024, morales2024}, high nitrogen-to-oxygen ratios \citep{Bunker2023, Cameron2023, Watanabe2024}, and high [OIII]$_{88\,\mu{\rm m}}$/[CII]$_{158\,\mu{\rm m}}$ ratios at $z>6$ \citep{Katz2022}.

It is also interesting to compare the predictions of the same model when a specific mechanism is implemented. For instance, the central panel of Fig.~\ref{fig:deltamodel} shows that the two feedback recipes implemented by \citet{Cantarella2025} give very similar results for faint galaxies ($M_{\rm UV}\sim-18$) but differ for the brightest ones ($M_{\rm UV}\sim-20.5$). Interestingly, the model that best fits the data is their implementation of the FFB scenario (with weaker stellar feedback at $z>4$), which is overall reasonably consistent with the original FFB implementation by \citet{Li2024}. This panel also includes the AFM model \citep{Ferrara2026UVRED}, in which a different type of feedback operates: when the radiation from star-forming regions exceeds a critical value, it sweeps away the dust, thus directly affecting the visibility of the objects rather than their star-formation properties.
On a more pessimistic note, the two versions of the \citet{Mauerhofer2025} model shown in Fig.~\ref{fig:deltamodel}, which implement either an enhanced star-formation efficiency (blue dashed lines) or an evolving IMF (blue solid lines), yield very similar predictions. This suggests that fitting the UVLF alone may not be enough to distinguish between competing explanations of its evolution.

Another crucial aspect is that none of the models consistently matches the data at all redshifts and luminosities. At least for some of the models that include free parameters, some redshift dependence in the implementation of their key physical processes is needed to explain the observations all the way from $z\sim10$ to $z\sim15$. Two examples, highlighted in the right panel of Fig.~\ref{fig:deltamodel}, are particularly informative. In the FFB models of \citet{LiDekel2024}, for instance, the star-formation efficiency (SFE) must evolve from $\epsilon_{\rm SFR}\sim0.2$ at $z\sim10$ to $\epsilon_{\rm SFR}\sim0.5$ at $z\sim15$ to fit the data. Similarly, in the models of \citet{Somerville2025}, which implement a strong increase in the SFE in high-density regions, with a fraction $f_{\rm d}=0.1$, 0.5, or 1 of the ISM in dense star-forming clouds, $f_{\rm d}$ must increase from 0.1 to at least 0.5 to remain close to the data. Although these models include a specific free parameter to describe these processes, they rely on physically motivated arguments suggesting that these effects become more important at high redshift, owing to an overall increase in the mean density of star-forming regions. Likewise, the AFM model \citep{Ferrara2026UVRED} explicitly includes a redshift-dependent process without additional free parameters.
These examples illustrate a more general conclusion: \textit{there is no reason to believe that the onset of specific physical processes occurs with  a sharp transition at a specific redshift}. It is much more likely that the transition from peculiar to more ordinary processes is gradual, evolving smoothly over the first 500~Myr. Likewise, there is no reason to believe that \textit{only one specific mechanism can explain the observed evolution}: a combination of different mechanisms may well operate at these early epochs and together conspire to produce the observed evolution. As noted by \citet{Somerville2025}, this explanation may be unsatisfying from the perspective of Occam's razor, but it is arguably the most plausible scenario. We refer the reader to \citet{Somerville2025} for an extended discussion of this aspect.

The overall conclusion is therefore that, although several physical processes can be invoked to explain the observed UVLF, \textit{the UVLF alone cannot distinguish among the different scenarios}, because very different physical mechanisms predict almost indistinguishable UVLFs (within the model and data uncertainties). We might summarize this section by saying that, while the question \enquote{can we produce enough bright galaxies at $z>10$?} has been answered in the affirmative by the latest models, the truly fundamental question \enquote{what physical mechanisms make them brighter and/or more abundant?} still awaits an answer.

\subsection{Cosmological implications of the JWST  UVLF}
\label{sec:theory-cosmo}

It should be clear at this point that several astrophysical mechanisms, involving only the baryonic component, are efficient enough to explain the number of galaxies at $z>10$. As a result, there is no strict need for alternative explanations that invoke modifications to the standard cosmological model. However, a number of works have proposed cosmological explanations for these early sources, or have explored the potential impact of non-standard cosmologies:\\
$\bullet$ The power spectrum of initial density perturbations can be modified due to the presence of primordial black holes (PBHs). This can lead to an enhancement in the abundance of massive halos at early epochs \citep{Zhang_PBH_MF2024,Matteri2025a}. However, this enhancement crucially depends on the poorly understood mass spectrum of PBHs \citep{dayal2026}, as well as on the similarly uncertain overall dark matter mass fraction contained in a PBH component \citep[e.g.,][]{Casanueva_PBH2025}\\
$\bullet$ An evolution of the dark energy equation of state \citep{menci2024} or the early emergence of dark energy \citep{shen2025} can change the growth rate of linear fluctuations and increase the number of massive haloes at high redshift, favoring the formation of bright/massive galaxies. An analogous effect arises in Wave Dark Matter models \citep{Zhang2026WDM}.\\
$\bullet$ Observations of the UVLF can provide us with  constraints on primordial magnetic fields \citep{fairbairn2026} or (non-standard) dark matter \citep{Gandolfi2022, maio2023, dayal2024, liu2024wdm}, at an epoch inaccessible by any other means.

While these modifications are certainly not \textit{needed} to explain the observed abundances, they may be motivated by other, more compelling cosmological observations, such as the latest DESI results \citep{Karim2025} or the Hubble tension \citep{Verde2019_tension, DESI2024_BAO, Riess2024_JWST, Wong2020_H0LiCOW, Freedman2026_CCHP}. We should therefore keep in mind that, if these modifications prove necessary to reconcile the findings of various cosmological probes, they will affect the expected number of early galaxies and hence the baryonic processes currently proposed to explain it.

\section{\enquote{Terra incognita}: Galaxies in the first 300 Myr}\label{sec:LF16}

While the observational picture up to $z\sim14.5$ is relatively well established, within the uncertainties described above, the search for galaxies beyond $z>15$ has so far produced only discordant UVLF estimates based on photometrically selected candidates, and the relatively few spectroscopic observations have failed to confirm any of them. In this section, we briefly summarize the status of the observations and the predictions of theoretical models, and conclude with a simplified analysis of future prospects.

\subsection{The search for galaxies  at \texorpdfstring{$z>15$}{z>15}}
\label{sec:z15_data}

At the time of writing, no galaxy has been spectroscopically confirmed beyond the current redshift record holder, MoM-z14 at $z=14.44$ \citep{Naidu2025}. This is not for lack of attempts. Several $z>15$ candidates have been observed with NIRSpec, and all have turned out to be low-redshift interlopers or remained inconclusive. The most striking example is CEERS-93316, which appeared to be a robust $z\simeq16$ candidate in the early NIRCam data \citep{Donnan2023} but turned out to lie at $z\simeq4.9$, owing to a rare combination of strong emission lines and a red continuum \citep{ArrabalHaro2023Nat}. A few other candidates, slightly fainter and somewhat less robust, have also been observed spectroscopically, yielding either further lower-redshift interlopers \citep{Castellano2025Z17} or inconclusive results \citep[as in the case of GOODS-N ID~1019411;][]{Hainline2026}. We therefore have to rely on purely photometric samples to derive the evolution of the UVLF at $z>15$, and the analyses conducted so far give strongly inconsistent results \citep{Castellano2025Z17, Kokorev2025, PerezGonzalez2025, Whitler2025, Weibel2026UVLF, McLeod2026, Gandolfi2026b}.

Before analyzing these papers in more detail, it is worth outlining the reasons that make this search particularly difficult. First, objects become progressively fainter, both because they are more distant and because they are expected to be intrinsically less luminous as we approach their epoch of formation. In addition, they shift into the redder NIRCam bands: $z\sim15$ galaxies are F200W dropouts and are therefore detectable only in the long-wavelength NIRCam channel (F277W and redder filters), where sensitivity starts to decline at $\lambda>4\,\mu$m owing to the increased background. Finally, at these red wavelengths and low apparent brightness, we become sensitive to families of interlopers that are very poorly known, such as low-luminosity quiescent or dusty star-forming galaxies at $z=4$--7. Although some of these may be interesting in their own right (e.g., Little Red Dots, or brown dwarfs that can mimic high-$z$ galaxies; \citealt{Holwerda2024, Bradac2026, Gandolfi2026a}), they make contamination even more difficult to estimate. In particular, these spectral classes may be underrepresented in the template libraries used in the SED fitting that yields $P(z)$, leading to the erroneous underweighting  of low-$z$ solutions. This combination of effects strongly affects the standard approach to selecting high-$z$ candidates, which relies on stringent criteria (e.g., high S/N thresholds and strict $P(z)$ cuts) to remove interlopers, combined with extensive simulations to correct for incompleteness; in short, an approach that favors purity and then corrects for incompleteness. With the existing data, this approach is more difficult and less productive at $z>15$, and differences in the specific procedures adopted (from data reduction to photometry, and from SED fitting to image simulations) make the various analyses difficult to compare. Ultimately, these difficulties stem from the lack of sufficiently deep and wide surveys, as we discuss below.

\begin{figure}[ht]
\centering
\includegraphics[width=0.95\linewidth,keepaspectratio]{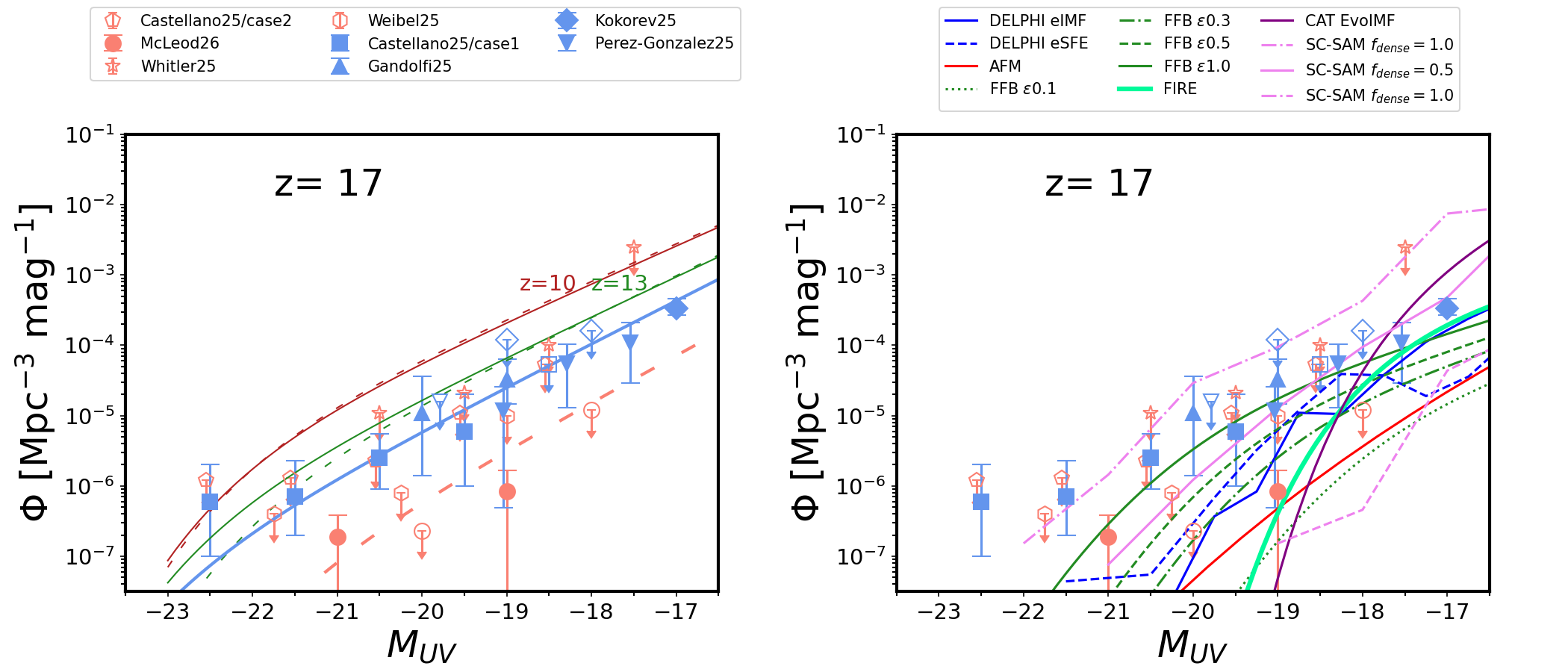}
\caption{UVLF at $z\sim17$. \textit{Left:} existing data at $z\sim17$, color-coded to distinguish the \enquote{z17-detections} papers (light blue points), which report detections of $z>15$ candidates (Case~1 of \citealt{Castellano2025Z17}; \citealt{Kokorev2025, PerezGonzalez2025, Gandolfi2026b}), from the \enquote{z17-upper limits} papers (salmon points), which primarily report upper limits (Case~2 of \citealt{Castellano2025Z17}; \citealt{McLeod2026, Weibel2026UVLF, Whitler2025}). Upper limits are shown as empty symbols. All detections refer to candidates without spectroscopic confirmation. The lines show the evolutionary UVLF fit of Sect.~\ref{sec:meta-analysis}, extended from $z=10$ to $z\sim17$ and computed separately for the two $z\sim17$ data sets: the solid light blue line includes the \enquote{z17-detections} data, and the dashed salmon line includes the \enquote{z17-upper limits} data. To place the $z\sim17$ results in context, the red and green lines show the corresponding best-fit UVLFs at $z=10$ and $z=13$, respectively; solid lines refer to the fit including the \enquote{z17-detections} data, and dashed lines to the fit including the \enquote{z17-upper limits} data. While the two fits diverge significantly at $z\sim17$, they are much closer at lower $z$, apart from some deviation at the bright end starting around $z\sim13$. \textit{Right:} comparison between the same $z\sim17$ observations and theoretical models. The data are those shown in the left panel; the lines (see legend) correspond to a subset of the models shown in Figs.~\ref{fig:models} and \ref{fig:deltamodel}.}
\label{fig:uvlf_z17}
\end{figure} 
%

This discussion helps interpret the discrepant results mentioned above. Some papers \citep{Castellano2025Z17, Kokorev2025, PerezGonzalez2025, Gandolfi2026b} have identified samples of \textit{plausible} candidates, which are typically close to the detection limit of the images, have colors consistent with a $z>15$ solution, and have a relatively small integrated $P(z)$ at lower $z$. All these papers explicitly discuss and quantify (via $P(z)$) the possibility that these objects are low-redshift interlopers. Assuming that the candidates are at $z>15$ and correcting for incompleteness, these studies derive positive measurements of the UVLF. Other papers, usually adopting more stringent criteria and different technical implementations, find no candidates \citep{Whitler2025, Weibel2026UVLF, McLeod2026}, or treat their formally detected candidates with caution \citep{McLeod2026}, and therefore yield only upper limits on the UVLF. \citet{Castellano2025Z17} also explicitly consider the possibility that all their candidates are interlopers, and compute the resulting UVLF in this scenario (hereafter Case~2, as opposed to Case~1, in which all candidates are assumed to be at $z>15$). In the following, we refer to the former set of results as \enquote{z17-detections} and to the latter as \enquote{z17-upper limits}.

These results are shown in the left panel of Fig.~\ref{fig:uvlf_z17}, where the two categories are plotted with different colors. The \enquote{z17-upper limits} studies point to a much faster evolution of the UVLF at $z>14.5$ than that implied by the \enquote{z17-detections} studies, which treat their candidates as plausible.
To quantify the impact of these two scenarios, we repeated the multi-parameter evolutionary fit of the UVLF over the extended redshift range $z=10$--17, keeping the two samples separate. Unlike the \enquote{Fiducial} model of Sect.~\ref{sec:fit}, we explicitly include the quadratic term $\log(\phi_2)$ as a free parameter. Using only the \enquote{z17-detections} studies (\citealt{Kokorev2025, PerezGonzalez2025, Gandolfi2026b}, and Case~1 of \citealt{Castellano2025Z17}), we find $\phi_2=0.007\pm0.004$, i.e., very small and consistent with zero within $2\sigma$, indicating that the UVLF would continue to evolve at a constant pace at these epochs. Using only the \enquote{z17-upper limits} studies (\citealt{Whitler2025, Weibel2026UVLF, McLeod2026}, and Case~2 of \citealt{Castellano2025Z17}), we instead find $\log\phi_2=-0.020\pm0.005$, indicating an evolution that accelerates with redshift. The two resulting predictions at $z\sim17$ are shown in Fig.~\ref{fig:uvlf_z17}. In both cases, the resulting evolution at $z<12.5$ is essentially unchanged. At bright magnitudes, however, the fit including the \enquote{z17-upper limits} data starts to underpredict the observed abundance of luminous galaxies at $z=13$--14.5. This suggests that the evolution is relatively steady up to $z\sim14.5$, and that, if the \enquote{z17-upper limits} results are correct, the quadratic form is too gentle to reproduce the sharp transition that would be required at $z>14.5$.

Finally, we note that the search for galaxies at very high redshift is by no means limited to F200W dropouts, which correspond to $z\lesssim21$ \citep{Castellano2025Z17,PerezGonzalez2025}. F277W and even F356W dropouts have also been searched for in existing data. While \citet{Castellano2025Z17} find no suitable candidates among $z\sim25$ F277W dropouts in the wide ASTRODEEP sample, which is roughly limited to $m_{\rm AB}\lesssim29.5$--30, \citet{PerezGonzalez2025} identify three candidates in the deeper MIDIS data, which reach $m_{\rm AB}\sim31.5$. Even more extreme F356W dropouts are being investigated, the most notable being \textit{Capotauro}, a relatively bright object ($m_{\rm F444W}\sim27.7$; \citealt{Gandolfi2026a}). For these objects, contamination by very cold Galactic brown dwarfs is a serious concern \citep{Gandolfi2026a}, as shown by follow-up observations of \textit{Capotauro}, which reveal a positional shift between epochs \citep{Liu2026}, and of two similar candidates, whose NIRSpec spectra are well fitted by Y-dwarf templates \citep{Bradac2026}.  For such bright candidates (at least by the standards of $z\sim30$ galaxies), the odds of being brown dwarfs remain very high. Nevertheless, interesting objects may still lurk within this class, including possible pair-instability supernovae \citep{Ferrara2026PISN,Jeon_PISN2026}, and we should remain open to these possibilities and continue to explore them. However, given the significant uncertainties in their selection, these $z>20$ candidates are not further discussed in the following.

\subsection{Prediction of theoretical models at \texorpdfstring{$z>15$}{z>15}}
\label{sec:z15_theory}

We now repeat the exercise of Sect.~\ref{sec:theory}, comparing the theoretical models developed in the JWST era with the scarce and uncertain observational evidence at these redshifts \citep[for a review of pre-JWST theoretical ideas, see][]{Bromm_Yoshida2011}. It is natural to expect that the new physical mechanisms invoked by the latest models to explain the high-redshift galaxy population become even more effective at higher redshift, which offers an opportunity to test these models. The right panel of Fig.~\ref{fig:uvlf_z17} compares the data shown in the left panel with most of the models of Fig.~\ref{fig:models}. We mostly restrict the comparison to models that already agree reasonably well with the data at lower $z$, with one exception: the model of \citet{Somerville2025} with $f_{\rm d}=1$, which  overestimates the data at lower $z$ (Fig.~\ref{fig:models}) and is included here as an extreme case.

With the caveat that the comparison is complicated by ambiguities in the data themselves (see Sect.~\ref{sec:z15_data}), a few results emerge:\\
\noindent
$\bullet$ The scatter among the models is larger than at $z=10$--14.5 (see Fig.~\ref{fig:models} for comparison). This is likely the result of the increasingly different impact of the novel physical mechanisms included in these models. However, other aspects of the model implementations may also play a role, including the details of the calculation of the halo mass function \citep[see][for a discussion]{Somerville2025} and the fact that the models are only weakly anchored by data at these redshifts.\\
$\bullet$  This increased scatter may be a valuable test to further constrain these models, once the $z>15$ observations are consolidated, possibly removing some of the degeneracies left after comparison with the data at lower $z$.\\
$\bullet$ There is a general tension between the models and the data, especially the data reporting detections (light blue points). Most models predict a faster evolution of the UVLF than implied by the \enquote{z17-detections} data and therefore fall short of the high densities derived from them. This behavior is a direct consequence of the evolution of DM halos in the standard $\Lambda$CDM model, whose mass function declines rapidly at these redshifts because the halos required to host these galaxies lie increasingly far in the exponential tail of the mass function.\\
$\bullet$ The tension between data and models is particularly evident for the brightest candidates, which are at least one order of magnitude more abundant than predicted by most models. Should even a fraction of them be confirmed, their existence would require significant revisions to our models. In contrast, the faintest candidates are easier to reproduce, and several models at least approach their density \citep[see also][]{Kokorev2025}.\\
$\bullet$ It is nevertheless possible to design models that match or even overpredict the most optimistic data. This, however, requires counterbalancing the decline in the abundance of massive halos by pushing the physical conditions to their limits, for instance by assuming that star formation is maximally efficient because of the absence of feedback \citep{LiDekel2024}, or because it occurs entirely in ultra-dense regions \citep{Somerville2025}. We note however that  these extreme models are disfavored by the data at lower redshift.

Overall, this comparison shows the importance of a solid census of galaxies at $z>15$, which current surveys have not yet achieved. Motivated by this, in the next section we explore the basic requirements for future surveys aimed at this goal.

\subsection{The UV Luminosity Density}

\begin{wrapfigure}{l}{0.5\textwidth}
\centering
\includegraphics[width=1\linewidth,keepaspectratio]{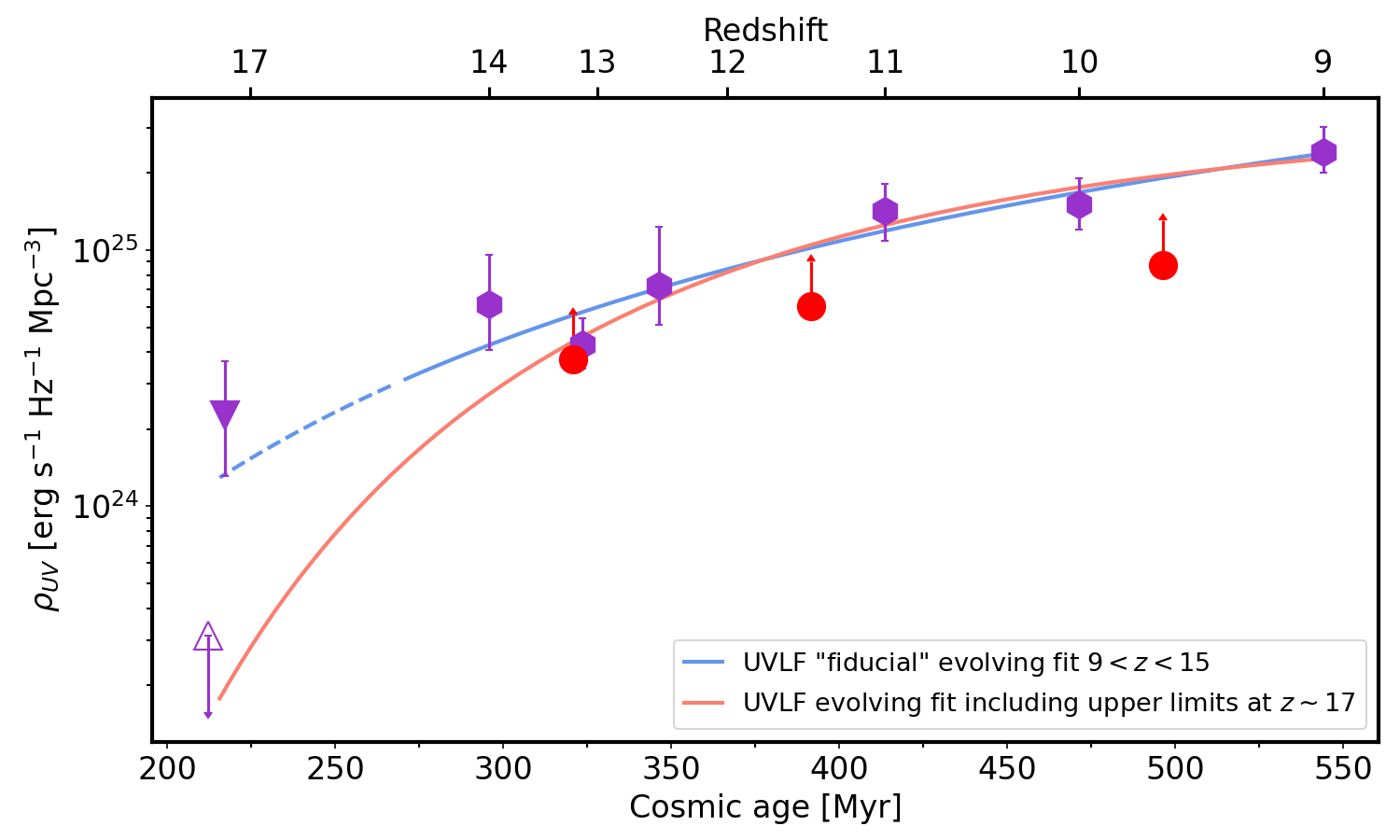}
\caption{Evolution of the UV luminosity density (UVLD), $\rho_{\rm UV}$, as a function of cosmic time since the Big Bang (lower x-axis) and redshift (upper x-axis). Individual points show the UVLD computed from the independent Schechter fits to the available UVLF data in each redshift bin. The two points at an age of $\sim230$~Myr ($z\sim17$) are computed from either the \enquote{z17-detections} studies (purple filled triangle) or the \enquote{z17-upper limits} studies (empty triangle); the latter should be regarded as an upper limit. The solid light blue line shows the UVLD from the \enquote{Fiducial} model, i.e., the evolving fit to all UVLF data from $z=9$ to $z=14.5$, and the dashed blue line is its simple extrapolation to an age of $\sim230$~Myr ($z\sim17$). The salmon line shows the seven-parameter evolutionary fit that also includes the \enquote{z17-upper limits} studies. All points and curves are integrated between $M_{\rm UV}=-24$ and $M_{\rm UV}=-17$. Red dots show the UVLD derived from the UVLF of spectroscopically confirmed galaxies only \citep{Harikane2025}, which, strictly speaking, should be regarded as lower limits.}
\label{fig:UVLD}
\end{wrapfigure}

The overall significance of these uncertainties for the global picture of galaxy evolution can be summarized in a familiar plot (Fig.~\ref{fig:UVLD}): the evolution of the UV luminosity density (UVLD), $\rho_{\rm UV}(z)=\int L\,\phi_{\rm UV}(L)\,dL$, where $\phi_{\rm UV}(L)$ is the UVLF. In practice, this quantity depends on the integration limits, in particular the faint limit: for a Schechter function with $\alpha<-2$, as found here ($\alpha\simeq-2.3$), the integral diverges at the faint end. For this reason, all values shown here are computed homogeneously between $M_{\rm UV}=-24$ and $M_{\rm UV}=-17$. To provide a self-consistent view of $\rho_{\rm UV}(z)$ fully anchored to the observed data, we first computed it from the evolutionary Schechter fit to the UVLF data described in Sect.~\ref{sec:fit}. The salmon line corresponds to the seven-parameter version of the fit, including the \enquote{z17-upper limits} data. The light blue line corresponds to the \enquote{Fiducial} model described above, extrapolated beyond its original redshift range; this extrapolation (dashed section) naturally matches the UVLD derived from the \enquote{z17-detections} studies. For comparison, the large points show the UVLD computed from the independent Schechter fits in each redshift bin, i.e., fitting all the data at a given redshift with a single, non-evolving Schechter function. We also show the UVLD derived from the UVLF of a purely spectroscopic sample \citep{Harikane2025}; these values should be regarded as strict lower limits. All data are plotted as a function of cosmic age (time since the Big Bang) to visualize the timescales involved.

The evolution of $\rho_{\rm UV}$ confirms a steady logarithmic decline in the density of galaxies with redshift up to $z\sim14.5$, well supported by the spectroscopic evidence. In particular, with as many as three relatively bright galaxies spectroscopically confirmed at $z\sim14$--14.5, the lower limit on $\rho_{\rm UV}$ is already high and very close to the value estimated from the combined UVLF data.

As described above, the two data sets at $z\sim17$ lead to completely different scenarios. While the fit matching the \enquote{z17-detections} studies would confirm a log-linear evolution of the UVLD up to $z\sim17$, the fit including the \enquote{z17-upper limits} determinations implies a drop of $\rho_{\rm UV}$ by about one order of magnitude between $z=14.5$ and $z\sim17$. Such a transition would have occurred over an interval of cosmic time that is very short in absolute terms ($\sim60$~Myr), but corresponds to about a quarter of the age of the Universe at that time.

This demonstrates the importance of confirming or disputing the small sample of candidates detected so far. Should a sizable fraction be confirmed, we would have to conclude that a smooth evolution of the UVLD continues up to these high redshifts, pushing the epoch of first galaxy assembly farther back in time. Otherwise, should the dearth of confirmed galaxies at $z>14.5$ be confirmed, we would be able to claim detection of the first major sharp transition in the evolution of this quantity, arguably one of the major transitions in the history of the Universe.  

\newpage

\subsection{Moving forward: prospects for the exploration of the "Terra Incognita"}
\label{sec:z15_future}

In this section, following \citet{Castellano2025Z17}, we predict the number of galaxies expected to be detected for various combinations of depth and area, using the models shown in the right panel of Fig.~\ref{fig:uvlf_z17} as a guide. Given how fast the UVLF is expected to evolve at $z>15$, we limit the comparison to models available over the relevant redshift range with a sampling of $\Delta z\leq1$, to ensure a smooth and accurate interpolation in redshift.

The number of objects (assuming 100\% completeness and purity) in a given area $A_\Omega$ can be computed with the standard formula
\begin{equation}
    N(z_1, z_2, m_{\rm F277W}, A_\Omega) = A_\Omega \int_{z_1}^{z_2} \frac{dV}{dz\,d\Omega} \left[ \int_{-24}^{M_{\rm max}(z,\,m_{\rm F277W})} \phi(M,z)\, dM \right] dz,
\label{Eq:counts}
\end{equation}
where $M_{\rm max}(z, m_{\rm F277W})$ is the rest-frame absolute UV magnitude corresponding to the limiting observed magnitude $m_{\rm F277W}$ at redshift $z$, $dV/(dz\,d\Omega)$ is the comoving volume element per unit redshift and solid angle, and $\phi(M,z)$ is the UVLF.

\begin{figure*}[ht]
\centering
\includegraphics[width=0.95\linewidth,keepaspectratio]{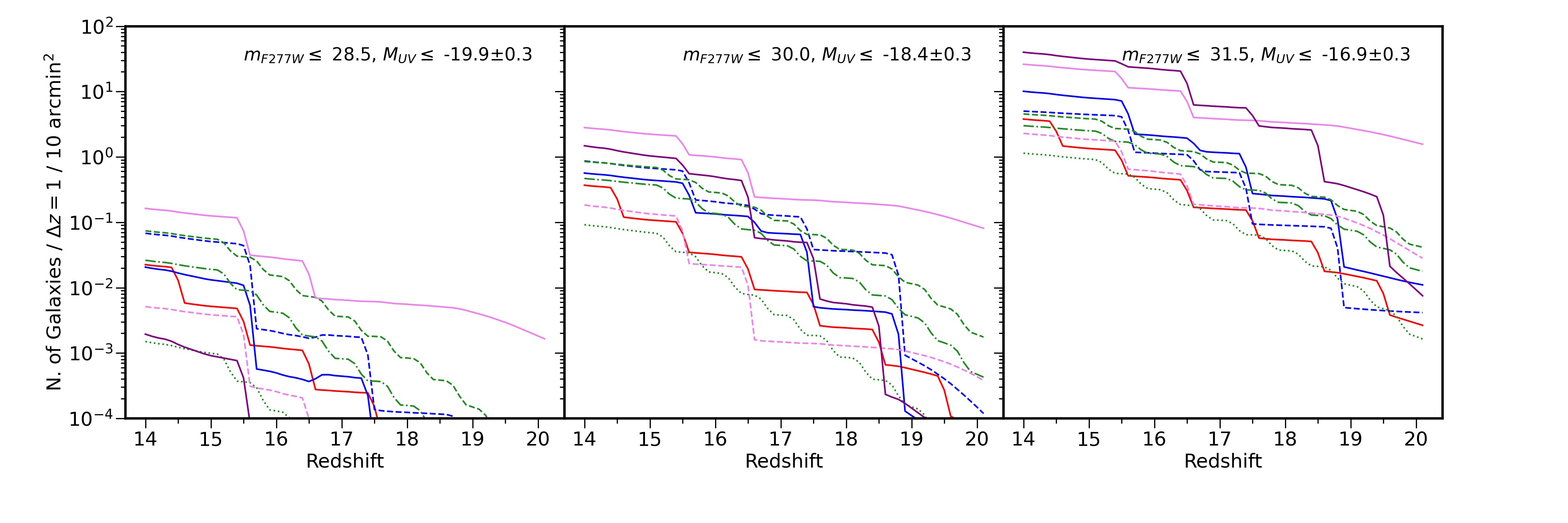}
\caption{Number of galaxies per unit redshift ($\Delta z=1$) brighter than limiting magnitudes of $m_{\rm F277W}=28.5$ (left), 30 (center), and 31.5 (right), as a function of redshift, in a field of 10~arcmin$^2$, according to a subset of the theoretical models discussed in this paper. Line styles and colors are as in Fig.~\ref{fig:models}. See the text for details.}
\label{fig:GalCounts}
\end{figure*}

The result of this calculation is shown in Fig.~\ref{fig:GalCounts}, which gives the number of galaxies predicted by these models as a function of the limiting magnitude in the NIRCam F277W band, over an area of 10~arcmin$^2$, very close to the NIRCam footprint of 9.7~arcmin$^2$. The counts in Fig.~\ref{fig:GalCounts} are normalized to a unit redshift interval, $\Delta z=1$. This calculation shows that, for all models, a limiting depth well beyond $m_{\rm F277W}=30$ is required for a realistic chance of detecting at least one galaxy at $z=16$--17 in a single ultra-deep NIRCam pointing. Alternatively, a much larger area must be surveyed.

\begin{figure*}[ht]
\centering
\includegraphics[width=0.95\linewidth,keepaspectratio]{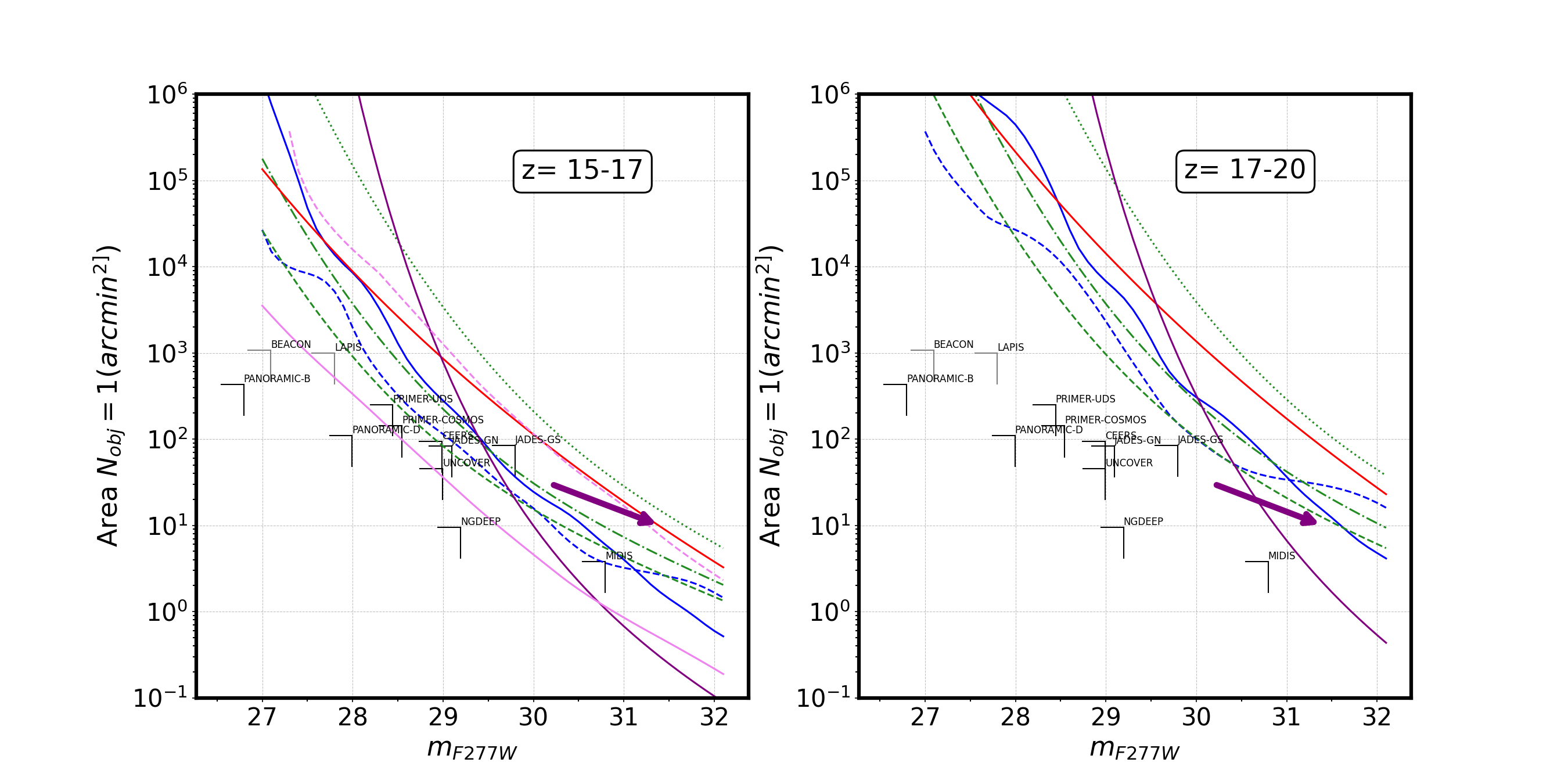}
\caption{Survey area (in arcmin$^{2}$) required for an expected number of \textit{one} galaxy  at $z=15$--17 (left panel) or $z=17$--20 (right panel), as a function of the limiting magnitude in the F277W detection band, according to a subset of the theoretical models discussed in this paper. Line styles and colors are as in Fig.~\ref{fig:models}. The areas and limiting magnitudes of major JWST surveys are shown in black for completed surveys and in gray for ongoing or planned ones. Their depths correspond to a robust detection of $z>15$ candidates ($S/N=8$ in the detection band, and an F200W depth sufficient to establish $m_{\rm F200W}-m_{\rm F277W}\geq1.5$ at $2\sigma$; see the text for details). The purple arrow shows the \enquote{lensing vector}, i.e., the gain in depth and the loss in sampled volume obtained by observing lensing clusters; its full length corresponds to a magnification $\mu=5$, i.e., a gain of 1.75~mag in depth and a factor of 5 reduction in area.}
\label{fig:areas}
\end{figure*}

This is extended in Fig.~\ref{fig:areas}, where we use Eq.~\ref{Eq:counts} to estimate the survey area required to detect \textit{one} object in a given redshift range $z_1\leq z\leq z_2$ as a function of the limiting magnitude $m_{\rm F277W}$. We perform this exercise in two redshift bins, $z=15$--17 and $z=17$--20, which broadly encompass the redshift range expected for F277W-detected F200W dropouts. Standard color selection with JWST filters cannot reliably discriminate between galaxies in the two bins, so a standard survey would collect candidates over the whole range $z=15$--20. However, given the fast predicted evolution of the UVLF in this range, it is instructive to show the two predictions separately.

Figure~\ref{fig:areas} also shows the observational limits of existing (black) and planned (gray) surveys. Among the completed surveys, we include CEERS \citep{Finkelstein2025}, UNCOVER \citep{Weaver2024,Wang2024uncover}, PRIMER-UDS and PRIMER-COSMOS \citep[][Magee et al., in prep.]{Dunlop2021}, JADES-GN and JADES-GS \citep{Eisenstein2023,Johnson2026_DR5}, NGDEEP \citep{Bagley2024}, MIDIS \citep[including the NGDEEP pointings;][]{PerezGonzalez2025}, and PANORAMIC \citep{Williams2025}. BEACON \citep{Morishita2025Beacon} is currently in progress, and LAPIS (GO~10790, PI S.~Tacchella) is scheduled to start soon. For most of these surveys, we adopt the typical depths estimated self-consistently in the ASTRODEEP catalog \citep{Merlin2024}. Several surveys use multi-tiered exposure schemes, so these depths should be taken as crude averages.

Crucially, for all these surveys except LAPIS, for which we adopt the limiting magnitude and area given in the approved proposal, we estimate the limiting magnitude taking full account of the S/N required for a robust detection of $z>15$ candidates. For simplicity, we require $S/N\geq8$ in the F277W detection band, and an F200W depth (at the $2\sigma$ level) sufficient to measure $m_{\rm F200W}-m_{\rm F277W}\geq1.5$, one of the typical color criteria used to select $z>15$ galaxies \citep{Castellano2025Z17}. In practice, given the depth of each survey in the various bands, we set the F277W detection limit to the brighter of \textit{a)} the magnitude corresponding to $S/N=8$ in F277W, and \textit{b)} $m_{\rm F200W}^{2\sigma}-1.5$, where $m_{\rm F200W}^{2\sigma}$ is the $2\sigma$ limiting magnitude in F200W. The S/N values are computed in small apertures ($0\farcs16$--$0\farcs2$, following the relevant papers) and corrected to total magnitudes assuming a point-like morphology.

First, this figure shows why current surveys fail to deliver robust results even at $z=15$--17. Even for the most optimistic models (among those consistent with the data), these surveys lie at best at a combination of depth and area that is barely sufficient to detect one object. Given the additional uncertainties and limitations of real observations compared with these somewhat simplistic estimates, it is not surprising that none of the candidates detected so far are not \enquote{bulletproof}.

Crucially, the figure also shows that, taking the median prediction of these models as a reference and aiming for a sample of at least 10 robust objects, a target that would allow useful constraints (at the $\sim 30\%$ error level) to be set, we need to plan a survey that is 1 dex higher than the median prediction of the models, markedly deeper and wider than those available today.

While these considerations hold for blank fields, deep surveys of lensing clusters provide a complementary approach. They cover sizable areas ($\gtrsim10\,{\rm arcmin}^2$) with magnifications $\mu\sim2$--5, where the study of the UVLF can be pushed to faint limits unattainable by unlensed surveys \citep{Castellano2016c,Atek2026,Chemerynska2026,Moreira_Goolsby2026}, and they also allow ultra-faint objects to be studied in detail, as discussed below. In Fig.~\ref{fig:areas}, the \enquote{lensing vector} shows the gain in depth (an increase of $2.5\log\mu$ in the effective limiting magnitude) and the corresponding loss in sampled volume (which scales as $1/\mu$) that lensed observations provide, and hence their impact on the probability of detecting $z>15$ objects. As shown in Fig.~\ref{fig:areas}, moderately lensed observations can increase the number of detected sources for a given observing time, at the cost of additional uncertainties in the sampled volume and in the intrinsic luminosity due to the uncertainty in $\mu$, which is, however, usually within 10--20\% at $\mu\lesssim5$. This effect, which holds for unresolved, background-limited sources, is a direct result of the steep slope of the UVLF predicted by all models shown in Fig.~\ref{fig:areas} and by the data, at least at lower $z$. At higher magnifications ($\mu\sim50$--100), these uncertainties increase significantly; nonetheless, such observations offer the unique opportunity to explore the internal structure of primordial galaxies (e.g., star-forming clumps) down to scales of a few parsecs, and to obtain spectra with a boosted S/N: since ${\rm S/N}\propto\mu$, the equivalent exposure time scales as $\mu^2$ \citep[e.g.,][]{Vanzella23_LAP1, Messa2026}.


\section{Discussion}
\label{sec:disc}

\subsection{Are we finding them all, and finding the right ones?}
\label{sec:disc_compl}
Although the observational picture appears relatively well settled, especially at $z<15$, considerable uncertainties remain in the observational chain that leads to the UVLF. The process involves many steps: \textit{a)} image processing (flat-fielding, identification of defects, calibration, astrometry, and coaddition); \textit{b)} noise estimation (production of the rms map), a crucial step for a proper statistical analysis; \textit{c)} source detection; \textit{d)} color measurement; \textit{e)} candidate selection, usually combining color-selection criteria with photometric redshifts and the associated probability of being truly at high $z$; and \textit{f)} estimation of completeness and selection effects through detailed simulations. Because high-redshift candidates are typically close to the noise level of the images, small differences in any of these steps can have a large effect. There is not yet a community-wide consensus on the specific procedures to adopt: the various recipes share many similarities, but also differ in many subtle details, and, as the saying goes, the devil is in the details.

Broadly speaking, the problem can be split into two steps: the selection of candidates and the calculation of the UVLF. For the first, the earliest analysis of the associated uncertainties in the JWST era was performed by \citet{Bouwens2023a}, who compared the candidates selected in the first three JWST fields. Their analysis revealed a remarkable difference even between two reductions of the same imaging data set, when the same selection criteria were applied to both. They found roughly the same number of $z>8$ candidates in the two reductions (36 vs.\ 39), but fewer than 50\% of the candidates were common to both samples. A comparison of the candidates selected by different authors on the same data set showed an even larger variance, up to a factor of 20 in the number of selected candidates, depending on how relaxed the selection criteria were. This very large variance was likely driven by the uncertainties of the first-generation pipelines, especially in the calibration of the early NIRCam data.

Clearly, as concluded by \cite{Bouwens2023a}, spectroscopy of complete and well-defined samples is ultimately the necessary tool to characterize efficiency, completeness and purity of the selection criteria. The first spectroscopic follow--up of a complete sample of $z>9$ galaxies was performed by \cite{Napolitano2025}, who follow--up a sample of 27 $z>9$ candidates (of varying quality and reliability) assembled in the A2744 field by \cite{Castellano2022b,Castellano2023b} with additional candidates selected by \cite{Atek2023, Harikane2023, McLeod2024}. \cite{Napolitano2025} have been able to confirm 7 galaxies at $z>9$ with only 1 lower redshift interloper, while the remaining candidates yielded inconclusive spectra, mostly owing to their faintness. Interestingly, two of the confirmed candidates were not originally selected by \cite{Castellano2023b} but were taken from other compilations. These results lent support to the notion that the LBG selection of $z>9$ is reliable, as already demonstrated by a wealth of spectroscopic confirmation on non-complete samples \citep[e.g.][]{Curtis-Lake2022, ArrabalHaro2023, ArrabalHaro2023Nat, Fujimoto2023, Fujimoto2023b, Hsiao2023, Hsiao2024, Naidu2025, Williams2023, Castellano2024, Roberts-Borsani2023, Roberts-Borsani2025Borg, Roberts-Borsani2026}.

Only a large, well-designed spectroscopic survey can put these initial results on a more solid footing. This is among the goals of the nearly completed CAPERS survey (PI M.~Dickinson). By design, CAPERS cast a wide net in selecting $z>8.5$ candidates, collecting potential targets from seven different analyses of the same data in the COSMOS, UDS, and EGS fields, and observing any candidate selected by at least one analysis. Donnan et al.\ (in prep.) analyze how the quality of the candidates correlates with the probability of being truly at high redshift. With 48 high-quality spectroscopic confirmations, they show that the success rate is, unsurprisingly, correlated with the number of independent analyses that identify a given candidate, and with the S/N and Lyman-break amplitude measured from the photometry. However, with as many as 135 sources identified as low-$z$ interlopers (about 74\% of the objects with a secure redshift, although many of them were marginal candidates) and another 144 objects without a reliable redshift, these results confirm that the selection of candidates is still subject to significant uncertainties and errors. 

The calculation of the UVLF from the available candidates is designed to correct for these effects through an extensive set of injection-recovery simulations. These typically reproduce all the steps involved in candidate selection to evaluate the completeness (and, in some cases, the contamination) of the sample, which is then used to correct the observed number of candidates and derive the final volume densities. These computationally intensive simulations are themselves subject to uncertainties and depend on the specific input parameters adopted. A crucial assumption, for instance, concerns the distribution of the spatial profiles of the candidates: at the faint limit of the images, where most candidates are found, more extended sources can easily escape detection. This is not merely a theoretical concern, as some of the highest-redshift spectroscopically confirmed galaxies are indeed extended \citep{Carniani2024a, Harikane2026}. An extended discussion of potential systematic errors in UVLF estimates is given by \citet{Leethochawalit2022c}.

One might wonder how these uncertainties affect the credibility of the UVLF results shown in the previous sections. In the redshift range $9<z<14.5$ these effects can easily explain the observed differences among the various results, which are at the level of 0.1--0.2~dex (Fig.~\ref{fig:uvlfresiduals}), together with statistical fluctuations and cosmic variance. The overall agreement among the various determinations at $z<15$ presented in the previous sections suggests that these effects do not dominate the results of the various analyses. They may, however, be more severe at the faint limits of the parent surveys, i.e., at the faint end of the UVLF and, for the shallower, wider-area surveys, also at bright magnitudes. In any case, there is room to improve the robustness of our measurements by making full use of the deepest data and of ongoing spectroscopic surveys.

The situation is more nuanced for the various UVLF determinations at $z>15$. Here, the candidates lie at the faintest limits of our observations, where these systematics are most severe. Given the CAPERS statistics shown above (Donnan et al., in prep.), it is not surprising that the few candidates followed up spectroscopically so far have not been confirmed at high redshift, and it is prudent to assume that many of the remaining high-$z$ candidates will turn out to be lower-redshift interlopers. However, past experience with HST and JWST candidates at lower $z$ has shown that plausible, though not robust, candidates have indeed been confirmed at high~$z$. We must therefore remain open to the possibility that at least some of the current or future plausible candidates are genuine high-$z$ sources, and push for their spectroscopic follow-up.

Finally, we note that at $z>7$ the contamination comes from low-redshift interlopers that mimic the properties of LBGs, i.e., they show a marked 4000\,\AA\ break at observed wavelengths $\lambda>1\,\mu$m and a relatively flat continuum redward of it or strong emission lines boosting the red bands. These $z>1.5$ galaxies are very different from the typical galaxies at such redshifts, which are blue, low-dust star-forming galaxies with a weaker 4000\,\AA\ break. These contaminants are either quiescent or dusty star-forming galaxies \citep[e.g.,][for the highest-redshift cases]{Castellano2025Z17}. Both classes are rare and poorly characterized. A full spectral characterization of these contaminants is therefore very important, not only because they are an interesting class of objects in their own right, but especially because our limited knowledge of them prevents us from properly accounting for their impact in the simulations of completeness and contamination used to measure the UVLF.

\subsection{Beyond the UVLF}
\label{sec:disc_prop}
It should be abundantly clear by now that the basic questions about the nature of $z>10$ galaxies cannot be addressed by the UVLF alone. Although the UVLF is useful for constraining some of their properties and their global evolution, a deeper investigation of their physical conditions is needed to determine which processes were at work at those early times. A number of papers have therefore investigated these objects in detail, combining multiwavelength data constraining their spectral energy distributions (SEDs), deep slit and, in some cases, integral-field spectroscopy, and ALMA observations. A full discussion of these studies is beyond the scope of this review: in this final section, we only sketch the main findings and open questions, with a necessarily incomplete bibliography.

Over the whole \enquote{epoch of reionization}, broadly corresponding to $z>6.5$ (with its final phase best traced by the visibility of Ly$\alpha$ emission; \citealt{Fontana2010, Stark2010, Mason2018, Jung2020, Napolitano2026Lya}), a number of studies have shown that the \textit{average} properties of galaxies change with redshift. For instance, these studies have established a trend toward lower average metallicities (from R23 measurements), higher excitation (from the intensity of the CIII] emission line), and younger stellar populations \citep{Trump2023, Tang2023, Nakajima2023, Curti2023, Roberts-Borsani2024, Tang2026survey}. Galaxies also show (mildly) bluer UV slopes \citep{Cullen2024, Roberts-Borsani2024, Dottorini2025, Felicioni2026, Hainline2026}, indicative of a progressively lower dust content and of younger and/or more ionizing stellar populations. These variations are, however, relatively modest, and other quantities, such as the specific star-formation rate \citep[sSFR;][]{Santini2026}, show little evolution in this redshift range. These findings point to a relatively gentle evolution in this epoch, consistent with the steady evolution of the UVLF over the same redshift range, and with the overall idea that the very first phase of galaxy assembly has not yet been observed, having likely occurred at $z>15$ and/or in objects too small to be detected today.

\bigskip
Within these global trends, these studies have also revealed an apparent bimodality in the galaxy population, as well as a wide \textit{diversity} within each of the two subpopulations.

Early spectroscopic follow-up observations of $z>10$ candidates revealed objects with prominent UV emission lines \citep[e.g.,][]{Castellano2024,Naidu2025,Napolitano2024b,Tang2026survey} within a larger population of objects with relatively featureless UV continua \citep{Curtis-Lake2022, Harikane2023, ArrabalHaro2023, Carniani2024a, Bunker2024}. Later studies \citep{Ono2025, Harikane2025, Roberts-Borsani2026} have shown that these high-ionization objects constitute a separate subpopulation. Strong emitters have barely resolved morphologies, consistent with half-light radii $r_e\leq100$~pc, as opposed to the global population, which appears spatially extended ($r_e>300$~pc), consistent with the evolutionary sequence of disk galaxies \citep{Harikane2025, Roberts-Borsani2026}. Strong emitters also show prominent N\,IV] emission lines implying a super-solar nitrogen abundance, often interpreted as indicative of the formation of stellar populations similar to those found in local globular clusters \citep[e.g.,][]{Charbonnel2023, DAntona2023, Senchyna2023, Marques-Chaves2024, Ji2025, Schaerer2026, Rai2026}. Finally, a number of objects have been identified as likely AGN hosts \citep{Maiolino2023, Kovacs2024, Napolitano2024b, ChavezOrtiz2025, Zhu2025, Fabian2026}, in most cases with $M_{\rm BH}/M_*$ ratios significantly higher than found at later times \citep[e.g.,][]{Maiolino2023, Napolitano2024b}.

Galaxies without strong UV emission lines also show a remarkable diversity. They span a broad range of UV slopes $\beta$, with the bulk of galaxies at $-2.6<\beta<-2.2$ \citep{Cullen2024, Roberts-Borsani2024, Dottorini2025, Felicioni2026}, typical of young stellar populations with little dust. Some of them, however, show very steep slopes, $\beta\sim-2.9$ \citep{Donnan2025,Cullen2025}, indicative not only of a negligible amount of dust but also of a high escape fraction of ionizing radiation, $f_{\rm esc}$ \citep{Marques-Chaves2026}, or of a low ISM density \citep{Katz2026}. Even more intriguingly, a few objects show $\beta$ values as large as $-1.5$, indicative of heavy dust attenuation \citep{Donnan2025, Mitsuhashi2025, Rodighiero2026, Parente2026}, although a spectrum dominated by nebular continuum emission is another viable explanation \citep{Mitsuhashi2025}. These red objects have been dubbed \enquote{Red Monsters}, to distinguish them from the more common \enquote{Blue Monsters}, i.e., objects with steep UV continua dominated by young stellar populations \citep[e.g.,][]{Aditya2026, Marques-Chaves2026VMS}.
Intriguingly, the AFM model described in Sect.~\ref{sec:theory-model} indeed predicts that the most active phase of star formation in a $z>10$ galaxy occurs in a dusty environment, until the radiation pressure from the star-forming regions exceeds a dust-modified Eddington limit and sweeps away the dust and gas. This model might naturally explain the transition from the \enquote{Red Monsters} (which have a limited visibility window, \citealt{Ferrara2024b,Santini2026}, and might be detected in far-IR observations, \citealt{Zavala2026}) to the more common \enquote{Blue Monsters}. Further evidence of diverse physical conditions comes from several studies reporting a variety of star-formation histories (SFHs) at high $z$ \citep[e.g.,][]{Dressler2024, Endsley2025, Santini2026, Harikane2026}, which also result in a spread in the mass-to-light ratios of the stellar populations \citep{Santini2023, Santini2026, Tang2026survey}. In this context, particular emphasis has been placed on an increased \enquote{burstiness} of the SFHs, traced by the H$\beta$ equivalent width \citep{Tang2025} or by the ratio between the recent and the long-term star-formation rate (e.g., ${\rm SFR}_{10}/{\rm SFR}_{100}$), which indicates a range of durations for the star-forming episodes \citep{Tang2025, Kokorev2025, Ciesla2024, Cole2025, Endsley2025, McClymont2025, Carvajal-Bohorquez2025}.

To complicate matters further, we remind the reader that most of this information is derived from the UV spectral range alone, since at $z>10$ the rest-frame optical spectrum is redshifted into the MIRI wavelength range and is hence more difficult to observe. Interestingly, the first spectroscopic observations with MIRI confirm the diversity of the $z>10$ galaxy population. While objects such as Maisie's Galaxy at $z=11.4$ \citep{Finkelstein2022Maise,ArrabalHaro2023} are confirmed to be typical star-forming galaxies evolving on the main sequence \citep{Larson2026}, strong UV line emitters show extreme ionization conditions and very high electron densities \citep{Zavala2025}. However, MIRI observations sometimes challenge interpretations based on the UV properties alone. The spectrum of GHZ1, a prototypical \enquote{normal} galaxy at $z\sim10$, for instance, shows an anomalous Balmer decrement that would require either extreme spectral properties or even the presence of a hidden AGN \citep{Perez-Diaz2026}. Conversely, the spectrum of another apparently \enquote{normal} galaxy, CEERS-588, shows no detectable H$\alpha$ or [O\,III]$\lambda5007$ emission \citep{Harikane2026}, indicating a prolonged star-formation history, possibly starting at $z>15$, followed by rapid quenching at least $\sim10$~Myr before the epoch of observation, in contrast to other $z>10$ galaxies. Other MIRI observations have challenged the interpretation of the high-$z$ AGN candidates GN-z11 \citep{CrespoGomez2026} and UNCOVER-26185 \citep{Alvarez-Marquez2026}, while recent NIRSpec data question the identification of UHZ1 as an AGN \citep{Zou2026}.

While most observations have so far targeted relatively bright galaxies, recent studies are starting to explore how the physical properties vary toward fainter galaxies. Early results \citep{Korber2026,Hsiao2026} suggest that low-luminosity galaxies are indeed more metal-poor than more luminous sources at the same redshift, but more observations are needed to explore this crucial regime. Again, programs such as SPURS (GO~9214, PIs Mason and Stark; \citealt{Chen2026SPURS}) and DAJE (GO~9645, PIs Treu, Fontana, and Roberts-Borsani) will mark a crucial step forward in this direction.

\bigskip
Finally, other observations suggest a variety and complexity of physical conditions \textit{within} individual galaxies. Evidence comes from individual extended sources such as GZ9p3 \citep{Bik2026}, and in more detail from the analysis of strongly lensed systems, most notably the \enquote{Cosmic Gems} arc at $z\sim9.6$ \citep{Messa2026} and the \enquote{Misty Moons} system at $z\sim11$ \citep{Nakane2025}, which exhibit a set of very dense star-forming clusters embedded in a fainter, more extended component. These clusters might be the observational counterparts of the very dense, compact star-forming regions implemented in some theoretical models \citep[e.g.,][]{Li2024, Somerville2025}. Combined ALMA$+$JWST observations \citep{Harikane2025} and advanced spectral synthesis models of strong UV line emitters \citep{Castellano2025Ion} also indicate diverse conditions in the ISM of these sources, with electron densities ranging from $\sim10$ to $10^6\,{\rm cm^{-3}}$ and ionization parameters from $\log U\sim-3$ to $\log U\sim0$. Overall, these observations suggest that the ISM of these objects is characterized by a combination of low-density, high-ionization clouds and, possibly, high-density, lower-ionization clouds. Spatially resolved integral-field and other spectroscopic observations are starting to reveal the complex dynamical interactions within the ISM, including the presence of a fully active baryon cycle \citep{Wang2026, Vasan2026, Zhu2026, Koller2026, Pascalau2026}.

\bigskip
It is difficult to say whether this wealth of discoveries helps answer the ultimate questions about the nature of these sources. As noted above, some results seem to align with specific prescriptions of individual models (e.g., \enquote{Red Monsters} for the AFM model, star-forming clumps for density-modulated models, and blue sources for the FFB mechanism), but a global picture has yet to emerge. Probably, this great diversity implies that a correspondingly large number of new observations is required to fully characterize the properties of these galaxies and to inform the models about the physical processes that govern their growth. Several models are already being tested against observables other than the UVLF, the stellar mass function and the mass--metallicity, luminosity--morphology, and SFR--stellar mass relations being obvious examples \citep{cueto2024, Cantarella2025, Vani2026}. Two additional lines of research can, in the long term, provide important independent constraints. The first is the analysis of the star-formation histories of individual galaxies prior to the epoch of observation, derived from detailed spectrophotometric analyses. These can indirectly probe epochs and mass ranges that are currently unexplored, and provide independent constraints for model validation. For instance, \citet{Donnan2025} and \citet{Santini2026} have shown that the main epoch of star formation of red objects with large $M/L$ can trace back to $z>15$, hinting at a vigorous phase of star formation beyond the epochs currently observed. The second is the study of large-scale clustering. Motivated by the growing availability of wide-field surveys, \citet{Weibel2026Clustering} and \citet{Nikolic2026} have shown that the large-scale distribution of galaxies can constrain the galaxy--halo connection (through the $M_{\rm UV}$--$M_{\rm halo}$ relation) and provide important clues to the star-formation process. Larger surveys, eventually combining data from JWST, Roman (most notably the ultra-deep RXDF; \citealt{Yan2026RXDF}), and Euclid, will open up this new field.

\section{Summary and concluding remarks}

This review focuses on our current understanding of the evolution of the UV luminosity function (UVLF) at $z>9$, as it emerges from four years of observations with the James Webb Space Telescope. We first briefly review (Sect.~\ref{sec:LF10}) why the first JWST discoveries were perceived to be in tension with the previous HST-based predictions for the evolution of galaxies at $z>10$. We then present (Sect.~\ref{sec:LF12}) a full compilation of the UVLF measurements obtained from blank-field and lensing-cluster JWST surveys. Taken together, these surveys (Fig.~\ref{fig:uvlfz9to14}) clearly show the global evolution of the UVLF, with the density of galaxies decreasing by a factor of $\sim7$  from $z=9$ to $z=14$. They also show that the UVLF is poorly sampled at both the bright end ($M_{\rm UV}<-22$ at $z=9$--12, and $M_{\rm UV}<-21$ at $z>12$) and the faint end ($M_{\rm UV}>-17$).
\bigskip

To put the analysis on a more quantitative footing, in Sect.~\ref{sec:meta-analysis} we perform an agnostic \enquote{meta-analysis} of the literature data, fitting them with a single redshift-dependent Schechter function from $z=9$ to $z=14.5$. We show in Fig.~\ref{fig:uvlfz9to14} that the data are reasonably well fitted by a single evolving function, which we dub the \enquote{Fiducial} model; the best-fit parameters are given in Table~\ref{tab:schechter_params}. We also performed a complementary analysis, dividing the sample into six redshift bins chosen to follow the redshift distribution of the data, and fitting an independent Schechter function in each bin. Even taking into account the uncertainties in the Schechter parameters and the degeneracies among them, we find that:\\
$\bullet$ the faint-end slope remains roughly constant at $\alpha\simeq-2.3$;\\
$\bullet$ the data are consistent with a uniform evolution of the galaxy density, that scales factor about $0.85$dex  between $z=9$ and $z=14$ if we adopt the \enquote{Fiducial} evolving fit. with no evidence of an accelerated evolution over the range $z=9$--14.5;\\
$\bullet$ The independent fits in each redshift bin also provide a good description of the data. Unlike the evolving fit, they show an evolution in both $\phi^*$ and $M^*_{\rm UV}$, with a milder evolution of $\phi^*$ compensated by $M^*_{\rm UV}$ becoming fainter with redshift. These differences are consistent with the uncertainties and degeneracies among the Schechter parameters (see Fig.~\ref{fig:uvlf_cornerplot}), and can most likely be attributed to the lack of sufficient data at the bright end. In any case, the two formulations give very similar descriptions in the range covered by the data.\\
$\bullet$ Because of the differences in the best-fit parameters, these fits should be used only as a convenient representation of the data, and should not be given a physical interpretation, for instance in terms of the evolutionary mode of the luminosity function (e.g., luminosity vs.\ density evolution) or of a differential evolution of bright and faint galaxies. Better coverage of the UVLF, especially at the bright end, is needed for more robust conclusions.\\
$\bullet$ Regardless of the specific fit, the density of bright galaxies at $z\sim14$ is surprisingly high compared with most determinations at $z\sim13$. In particular, the most recent large-area studies at $z\sim13$ \citep{McLeod2026,Weibel2026UVLF} yield galaxy densities that are lower than the bulk of other determinations at $z\sim13$ and even at $z\sim14$, including those based on spectroscopic samples \citep{Harikane2024};\\
$\bullet$ These discrepancies are part of a more general pattern of fluctuations between the data and the smooth evolving fit at various redshifts (Fig.~\ref{fig:uvlfresiduals}), at the level of 0.1--0.2~dex, indicating that systematics in the data and/or the effects of large-scale structure are not yet averaged out in the existing data.
\bigskip

Using the \enquote{Fiducial} model as a smooth representation of the data, in Sect.~\ref{sec:theory} we present an extended comparison between the UVLF data and a suite of theoretical \textit{ab initio} models of galaxy evolution, focusing on those that have recently included specific physical processes to better match the observations. We show (Figs.~\ref{fig:models} and \ref{fig:deltamodel}) that:\\
$\bullet$ no single model matches the observed density of galaxies at all $M_{\rm UV}$ and redshifts in a stringent statistical sense, i.e., within the $\sim0.2$~dex accuracy of the data;\\
$\bullet$ however, thanks to the inclusion of new physical processes motivated by the different physical conditions expected in the ISM at high redshift, these models are generally able to match (and often even exceed) the observed galaxy densities up to $z=14.5$. The observed high density of galaxies up to $z=14.5$ therefore does not, per se, represent a challenge to these models;\\
$\bullet$ the UVLF data alone cannot discriminate among the different processes proposed, as models with different physical ingredients reproduce the UVLF with similar quality.

We conclude that, while the question \enquote{can we produce enough bright galaxies at $z>10$?} raised by the earliest JWST observations has been answered in the affirmative by the inclusion of new, physically motivated mechanisms, the truly fundamental question \enquote{which of these mechanisms, or which combination of them, makes galaxies brighter and/or more abundant?} still awaits an answer. For the reasons described above, this question is unlikely to be settled by more refined UVLF measurements alone. Substantial progress in the field can only come from an in-depth analysis of the physical conditions in early galaxies, combining a variety of probes.
\bigskip

Finally, in Sect.~\ref{sec:z15_data} we summarize the results of the so far unsuccessful searches for galaxies at $z>15$. We show that:\\
$\bullet$ although several plausible candidates have been identified, none has been spectroscopically confirmed or can be considered \enquote{bulletproof}, as they lie at the faintest limits of existing surveys;\\
$\bullet$ depending on how the selection criteria are applied, different studies obtain strongly discrepant estimates of the evolution of the UVLF (Fig.~\ref{fig:uvlf_z17}) and of the UVLD (Fig.~\ref{fig:UVLD}) up to $z\sim17$;\\
$\bullet$ if the studies reporting \enquote{plausible} candidates (\citealt{Kokorev2025, PerezGonzalez2025, Gandolfi2026b}, and Case~1 of \citealt{Castellano2025Z17}), which we refer to as \enquote{z17-detections}, are correct, the decline of the UVLF follows the gentle trend found at $z\lesssim14.5$.

$\bullet$ conversely, if the studies adopting more stringent criteria (\citealt{McLeod2026, Weibel2026UVLF, Whitler2025}, and Case~2 of \citealt{Castellano2025Z17}), which we refer to as \enquote{z17-upper limits}, are correct, the derived upper limits imply a very sharp decrease in the galaxy density beyond $z\sim14.5$, suggesting that the UV-bright galaxy population emerged very rapidly just before that epoch (Figs.~\ref{fig:uvlf_z17} and \ref{fig:UVLD});\\
$\bullet$ in any case, the evolution of the UVLF at $z>15$ provides a very stringent test of existing theoretical models (Sect.~\ref{sec:z15_theory}). The models diverge at these redshifts, even those that are degenerate at lower $z$. In addition, only extreme physical conditions seem able to reproduce the densities derived from the \enquote{z17-detections} studies. The difficulty that theoretical models have in reproducing these measurements at $z\sim17$ stems largely from the evolution of the DM halo mass function, which declines steeply at these redshifts. For both reasons, a firm determination of the UVLF at these redshifts holds the promise of a remarkable advance in our understanding of the early Universe.

\bigskip

To resolve the observational uncertainties, the usual conclusion that any astronomer would draw is a plea for more observations: larger, deeper, and more diverse. These are indeed needed: a combination of new deep (especially in lensing fields) and wide surveys is required to fully sample the UVLF at the faint ($M_{\rm UV}>-17$) and bright ($M_{\rm UV}<-22$) ends, which are very poorly sampled, especially at $z>12$. In addition, systematic spectroscopic follow-up is needed to secure the redshifts, especially at the faint end. While this is true at any redshift, it is of the utmost importance to enter the \textit{terra incognita} beyond the current spectroscopic redshift record of $z=14.44$, where we need to confirm or refute the nature of the current candidates through ultra-deep spectroscopy, and to extend the search for photometric candidates with dedicated surveys such as LAPIS (GO~10790, PI S.~Tacchella).

After four years of JWST observations, there is growing evidence that gathering the data needed to address these points may require an effort beyond the scale of current JWST programs. To provide a first estimate of the required combination of depth and area, in Sect.~\ref{sec:z15_future} (Figs.~\ref{fig:GalCounts} and \ref{fig:areas}) we compute the expected galaxy counts at $z>15$ predicted by current theoretical models.

\bigskip
New data, however, are not the only ingredient we need. After four years of JWST observations, the \enquote{low-hanging fruit} has been harvested, and the transformative power of its observations is likely to diminish. No future space-based facility currently under construction or in design will surpass JWST's infrared sensitivity beyond $2\mu$m, and spectroscopic follow-up with ELT-class instrumentation will not resolve all outstanding questions. As discussed in Sect.~\ref{sec:disc_compl}, differences in data-processing procedures introduce additional uncertainties that affect the robustness of, and the consensus on, the derived results. A coordinated, community-wide effort to compare and consolidate the techniques used to derive the UVLF from imaging surveys is therefore needed, alongside targeted observational campaigns.

\smallskip

This combined endeavor will constitute the ultimate legacy of a unique and irreplaceable facility---and will, we hope, unveil the birth of the first galaxies.
\bigskip

\section*{Acknowledgments}
AF warmly thanks the organizers of the meeting for the kind invitation and the very productive and lively workshop. The authors thank the Sexten Center for Astrophysics Riccardo Giacconi for its hospitality and inspiring environment, which stimulated many discussions that led to this work. AF acknowledges that this work was carried out within the COSMOWEBB project “A new era for cosmology: exploiting the JWST revolution”, funded by the Italian National Institute for Astrophysics (INAF) through the PRORIS programme (PROgramma di RIcerca Spaziale di base), CUP C83C26000300005.
PD  acknowledges support from an NSERC discovery grant (RGPIN-2025-06182).
PS acknowledges support from the RF2024 Large Grant "UNDUST: UNveiling the Dawn of the Universe with JWST". MC acknowledges INAF GO Grant 2024 ”Revealing the nature of bright
galaxies at cosmic dawn with deep JWST spectroscopy”. PGP-G acknowledges support from grant PID2022-139567NB-I00 funded by Spanish Ministerio de Ciencia, Innovaci\'on y Universidades MCIU/AEI/10.13039/501100011033,
FEDER {\it Una manera de hacer Europa}.
LC acknowledges support from the French government under the France 2030 investment plan, as part of the Initiative d’Excellence d’Aix-Marseille Université – A*MIDEX AMX-22-RE-AB-101.
AJB acknowledges funding from the ``FirstGalaxies" Advanced Grant from the European Research Council (ERC) under the European Union’s Horizon 2020 research and innovation programme (Grant agreement No. 789056)
FDE acknowledges support by the Science and Technology Facilities Council (STFC), by the ERC through Advanced Grant 695671 ``QUENCH'', and by the UKRI Frontier Research grant RISEandFALL.

\bigskip
All the UVLF data collected and used in this paper are available upon request to the first author.

\bigskip

We acknowledge the use of Claude (Anthropic; model Claude Opus 5.5) for assistance with coding, literature searches, and grammar checking. All outputs were reviewed and verified by the authors, who take full responsibility for the content of this paper.

\bibliographystyle{mn2e} 
\bibliography{biblio3.bib}

\begin{appendix}
\label{sec:app1}

\begin{wrapfigure}{l}{0.57\textwidth}
\centering
\includegraphics[width=0.8\linewidth,keepaspectratio]{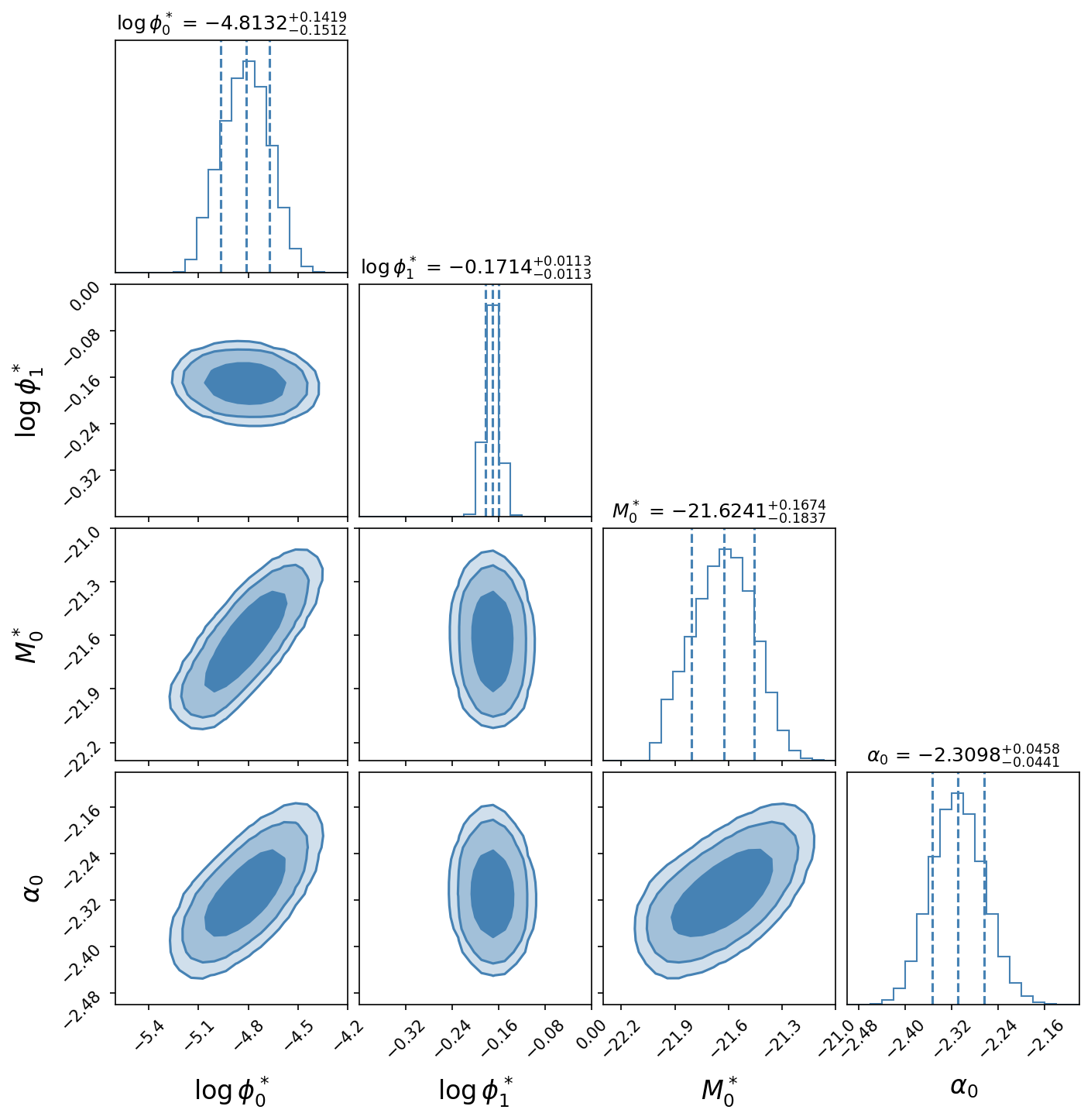}
\caption{Error analysis of the four free parameters of the \enquote{evolutionary} fit to the $z-9-14.5$ UVLF with a single Schechter Function. See text for further details.}
\label{fig:uvlf_cornerplot}
\end{wrapfigure} 

Here we describe how we performed the fit to the UVLF over the range $8.5<z<14.5$. For this calculation, we included the available UVLF determinations at $8.5<z<14.5$ \citep{\UVLFfit}, with the associated errors as quoted by the authors. We did not use the data of \citet{Harikane2025}, since they are, strictly speaking, lower limits. Each data point was assigned the central redshift of the relevant bin or, where available, the median redshift quoted by the authors.

We started from the Schechter parameterization discussed in the text, with three free parameters (the normalization $\phi(z)$, the faint-end slope $\alpha(z)$, and the characteristic magnitude $M^*(z)$), each evolving from $z=9$. We initially allowed all three parameters to vary as a function of $(z-9)$, as shown in Eq.~\ref{Eq:EvolvingSchechter}. In particular, $\alpha(z)$ and $M^*(z)$ evolve linearly, $M^*(z)=M^*_0+M^*_1\,(z-9)$ and $\alpha(z)=\alpha_0+\alpha_1\,(z-9)$, while $\phi(z)$ follows a quadratic form, $\log\phi(z)=\log(\phi_0)+\log(\phi_1)\,(z-9)+\log(\phi_2)\,(z-9)^2$. In this form, $\alpha_0$, $\log(\phi_0)$, and $M^*_0$ are the values of the parameters at $z=9$. The fit was performed with the \textit{emcee} Python package, and we verified that the number of walkers (63) and steps (200,000) is sufficient to yield stable results for different initial configurations.

Upper limits in the UVLF are handled in the likelihood with a soft-penalty approach. For detections, the log-likelihood contribution is the standard asymmetric Gaussian term. For data points flagged as upper limits (those with non-positive error bars), we instead apply a soft penalty, i.e., they contribute to the likelihood only when the model exceeds the upper limit. We also applied an error floor of 50\% to all individual measurements, to prevent measurements with very small errors from dominating the fit; this choice does not have a significant impact on the final fit. Finally, we excluded from the fit the faintest data points of \citet{Chemerynska2026} and \citet{Finkelstein2024}, for which the incompleteness correction exceeds 100\%.

As discussed in the main text, after several tests we progressively fixed to zero all the parameters that were consistent with zero within the errors, to avoid an over-parameterized fit. We eventually converged on a four-parameter form for our \enquote{Fiducial} run, setting to zero both the quadratic term of the normalization, $\log(\phi_2)\equiv0$, and the linear terms of the other two parameters, $M^*_1\equiv\alpha_1\equiv0$. Reducing the number of free parameters does not increase the reduced $\chi^2$: we obtain $\chi^2_{\rm red}\simeq1.3$ for the \enquote{Fiducial} fit and $\simeq1.35$ for the seven-parameter fit. For about 200 degrees of freedom, this excess over unity corresponds formally to a $3.3\sigma$ deviation from the value expected if the reported errors were purely statistical, indicating that additional sources of error are present. As shown in Fig.~\ref{fig:uvlfresiduals}, systematic errors indeed contribute significantly to the residuals.

Table~\ref{tab:schechter_params} lists the best-fit parameters of our four-parameter \enquote{Fiducial} fit, and Fig.~\ref{fig:uvlf_cornerplot} shows their posterior distributions. The corner plot shows the well-known correlation between $M^*_0$ and $\log(\phi_0)$.

As mentioned in the main text, we also performed a complementary analysis, dividing the whole redshift interval into six redshift bins and fitting the data in each of them separately. The bins are approximately one unit in redshift wide, but were adjusted to reflect the actual redshift distribution of the data (which shows a marked gap at $z\sim11.5$--12, as discussed in the main text). In each bin, we fitted the available data with a non-evolving Schechter function, leaving free only its three parameters $\log\phi^*$, $M^*_{\rm UV}$, and $\alpha$. The results are also listed in Table~\ref{tab:schechter_params}. As discussed in the main text, $M^*_{\rm UV}$ and $\log\phi^*$ show a slightly different evolution from that of the \enquote{Fiducial} model, with $M^*_{\rm UV}$ becoming fainter with redshift and a correspondingly milder evolution of $\log\phi^*$. This behavior is consistent with the known degeneracy between $M^*_{\rm UV}$ and $\log\phi^*$ (see Fig.~\ref{fig:uvlf_cornerplot}). However, the resulting UVLFs are practically indistinguishable from the \enquote{Fiducial} evolving model, at least where data exist, as shown in Fig.~\ref{fig:uvlfz9to14}.

\begin{table*}
\centering
\label{tab:schechter_params}
\renewcommand{\arraystretch}{1.8}
\begin{tabular}{|l|c|cccccc|}
\hline\hline
Parameter & Evolving fit & $z=9$ & $z=10$ & $z=11$ & $z=12$ & $z=13.12$ & $z=14$ \\
\hline
$\log\phi^*_0$ & $-4.81^{+0.14}_{-0.15}$ & $-4.82^{+0.15}_{-0.12}$ & $-4.94^{+0.18}_{-0.16}$ & $-4.84^{+0.15}_{-0.16}$ & $-5.06^{+0.12}_{-0.10}$ & $-5.17^{+0.04}_{-0.02}$ & $-5.08^{+0.10}_{-0.08}$ \\
$\log\phi^*_1$ & $-0.17^{+0.01}_{-0.01}$ & \multicolumn{1}{c}{Fixed $= 0$} & \multicolumn{1}{c}{Fixed $= 0$} & \multicolumn{1}{c}{Fixed $= 0$} & \multicolumn{1}{c}{Fixed $= 0$} & \multicolumn{1}{c}{Fixed $= 0$} & \multicolumn{1}{c|}{Fixed $= 0$} \\
$M^*_0$ & $-21.62^{+0.17}_{-0.18}$ & $-21.70^{+0.18}_{-0.14}$ & $-21.50^{+0.25}_{-0.23}$ & $-21.22^{+0.17}_{-0.21}$ & $-21.16^{+0.17}_{-0.19}$ & $-20.95^{+0.04}_{-0.06}$ & $-21.03^{+0.09}_{-0.14}$ \\
$\alpha_0$ & $-2.31^{+0.05}_{-0.04}$ & $-2.25^{+0.06}_{-0.05}$ & $-2.28^{+0.05}_{-0.06}$ & $-2.33^{+0.07}_{-0.05}$ & $-2.27^{+0.05}_{-0.07}$ & $-2.28^{+0.05}_{-0.06}$ & $-2.30^{+0.07}_{-0.07}$ \\
\hline
\end{tabular}
\caption{Best-fit Schechter function parameters obtained by fitting the literature data used in the text.  The first column reports the names of the 4 free parameter  of our fiducial model, which uses a linearly evolving Schechter function to describe the evolution of the UV LF from $z=9$ to $z=14.5$.   See Eq.\ref{Eq:EvolvingSchechter} and text for details. Column 2 reports the best-fit values of the 4-parameter fit with their errors.
Columns 3-8 show the parameters of fit obtained within individual redshifts bins. Here the fit is performed with only zero-order parameters free.}
\end{table*}

\begin{figure}[ht]
\centering
\includegraphics[width=0.95\linewidth,keepaspectratio]{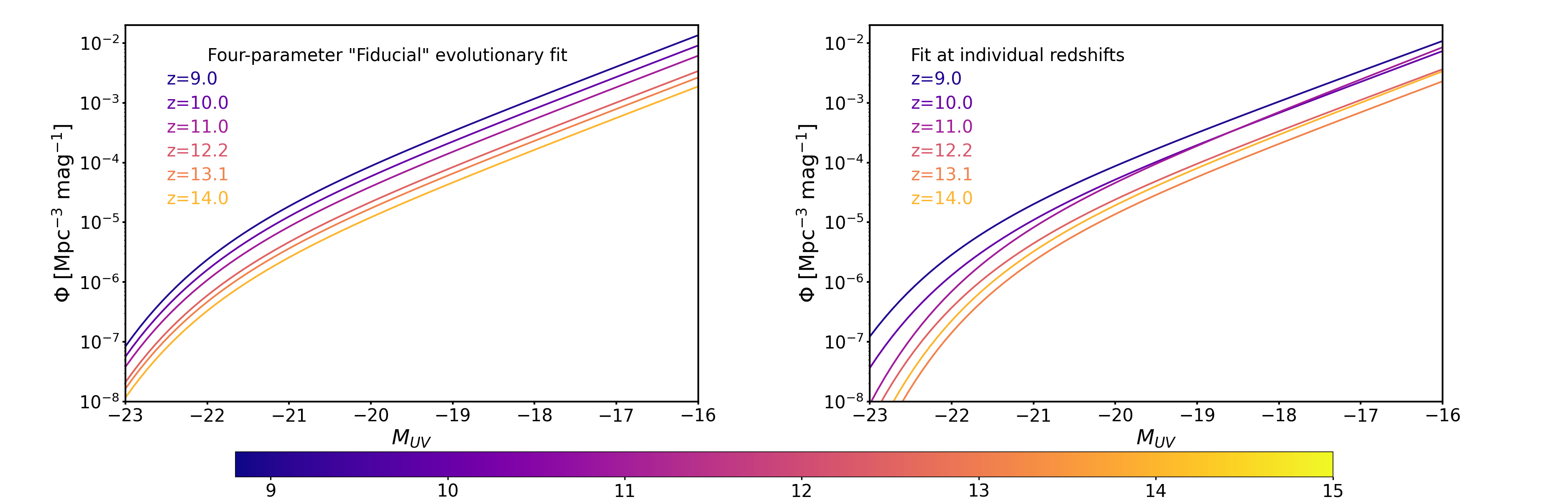}
\caption{ Redshift evolution of the \enquote{Fiducial} four-parameter evolutionary fit (leftmost panel) of the UVLF from $z=9$ to $z=14.5$, compared to the fit to individual redshift ranges (rightmost panel). Curves are colored according to their redshift as shown by the legend and by the color bar at the bottom. The rightmost panel shows how the $z\sim14$ UVLF sits above the $z=13.1$ one, an example of the uncertainties still affecting the estimate of the UVLF. Data used to estimate the fit are reported in Fig.~\ref{fig:uvlfz9to14}. See text for full details.}
\label{fig:evolutionari_fit}
\end{figure}

Figure~\ref{fig:evolutionari_fit} compares the redshift evolution of the to Schechter fits. The left panel shows the four-parameter \enquote{Fiducial} evolving fit, and the right panel shows the independent fits in the corresponding redshift bins. This comparison confirms the overall smooth evolution, but also shows that the $z\sim14$ single-bin UVLF lies \textit{above} not only the level expected from the \enquote{Fiducial} model, but even the $z=13.1$ single-bin UVLF. This result is driven by the high density of bright galaxies at $z\sim14$, several of which are spectroscopically confirmed, and would imply that the evolution from $z=13$ to $z=14$ is milder than assumed by the linear evolution of the \enquote{Fiducial} fit, or even reversed. Given the small size of the current sample, it is too early to draw any firm conclusion from this result: wider surveys are needed to establish whether this trend is real or an effect of the limited sample.


\end{appendix}

\end{document}